\documentclass[sigconf]{acmart}

\setcopyright{none}

\renewcommand\footnotetextcopyrightpermission[1]{}

\usepackage{graphicx}
\usepackage[bottom]{footmisc}
\usepackage{subfig}
\usepackage{epigraph}
\usepackage{multirow}
\usepackage{array}
\usepackage{amsmath}
\usepackage{amsfonts}
\usepackage{algorithm,algorithmic}
\usepackage{float}
\usepackage{balance}

\begin{document}
\title{Deep Models Under the GAN: Information Leakage from Collaborative Deep Learning} 

\author{Briland Hitaj}
\authornote{The author is also a PhD student at University of Rome - La Sapienza}
\affiliation{%
  \institution{Stevens Institute of Technology}
}
\email{bhitaj@stevens.edu}

\author{Giuseppe Ateniese}
\affiliation{%
  \institution{Stevens Institute of Technology}
}
\email{gatenies@stevens.edu}

\author{Fernando Perez-Cruz}
\affiliation{%
  \institution{Stevens Institute of Technology}
}
\email{fperezcr@stevens.edu}

\renewcommand{\shortauthors}{B. Hitaj et al.}

\begin{abstract}

Deep Learning has recently become hugely popular in machine learning for its ability to solve end-to-end learning systems, in which the features and the classifiers are learned simultaneously, providing significant improvements in classification accuracy in the presence of highly-structured and large databases.  

Its success is due to a combination of recent algorithmic breakthroughs, increasingly powerful computers, and access to significant amounts of data. 

Researchers have also considered privacy implications of deep learning. Models are typically trained in a centralized manner with all the data being processed by the same training algorithm. If the data is a collection of users' private data, including habits, personal pictures, geographical positions, interests, and more, the centralized server will have access to sensitive information that could potentially be mishandled. To tackle this problem, collaborative deep learning models have recently been proposed where parties locally train their deep learning structures and only share a subset of the parameters in the attempt to keep their respective training sets private. Parameters can also be obfuscated via differential privacy (DP) to make information extraction even more challenging, as proposed by Shokri and Shmatikov at CCS'15.

Unfortunately, we show that any privacy-preserving collaborative deep learning is susceptible to a powerful attack that we devise in this paper. In particular, we show that a \emph{distributed, federated, or decentralized deep learning} approach is fundamentally broken and does not protect the training sets of honest participants.   
The attack we developed exploits the real-time nature of the learning process that allows the adversary to train a Generative Adversarial Network (GAN) that generates prototypical samples of the targeted training set that was meant to be private (the samples generated by the GAN are intended to come from the same distribution as the training data). Interestingly, we show that record-level differential privacy applied to the shared parameters of the model, as suggested in previous work, is ineffective (i.e., record-level DP is not designed to address our attack). 
\end{abstract}

\keywords{Collaborative learning; Security; Privacy; Deep learning}

\maketitle

\epigraph{It's not who has the best algorithm that wins.\\
It's who has the most data.}{\textit{Andrew Ng \\ Self-taught Learning}}

\section{Introduction}
\label{introduction}

\begin{figure*}
\begin{tabular}{cc}
\subfloat[Centralized Learning]{\includegraphics[width = 3.5in]{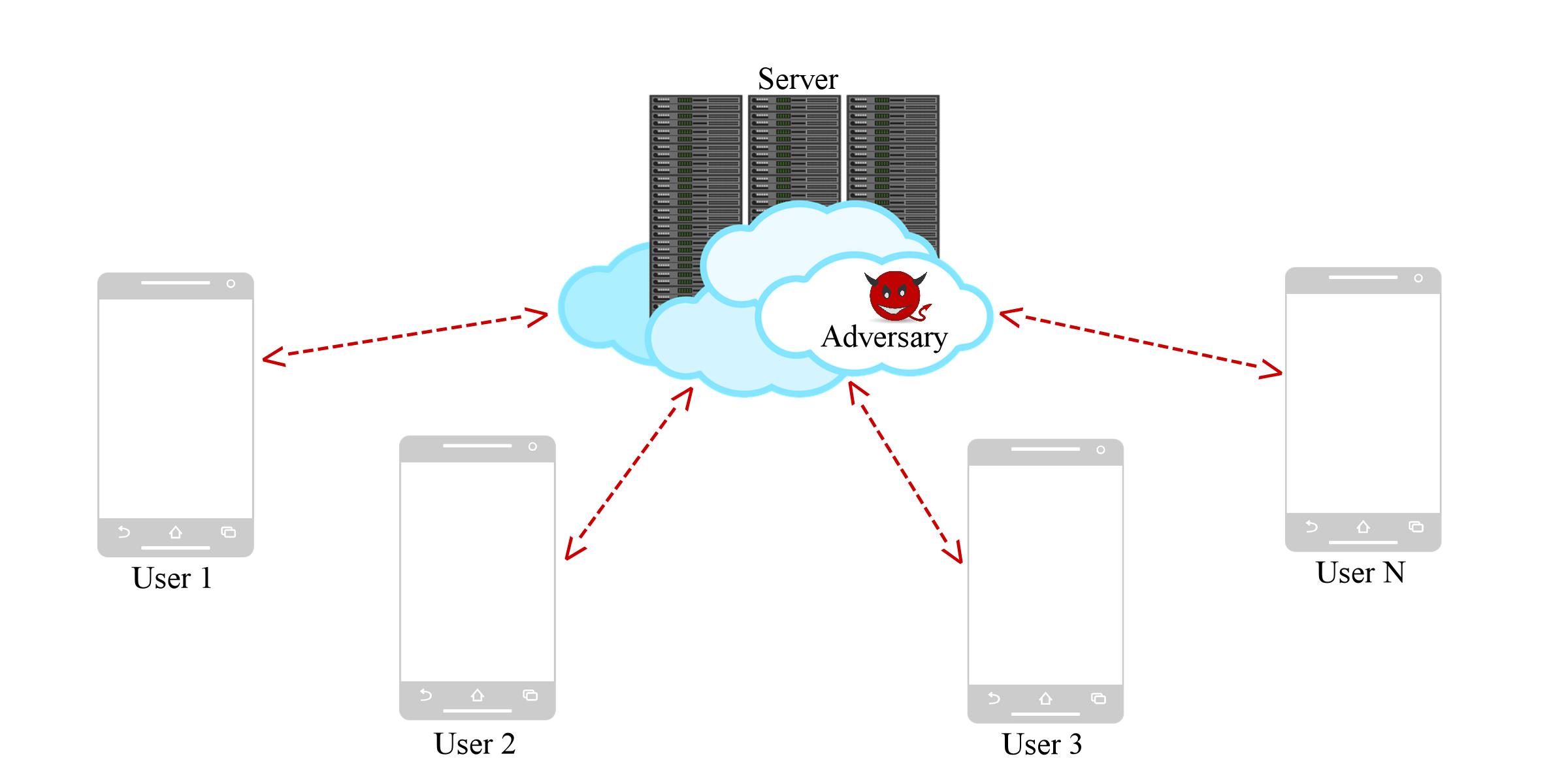}} &
\subfloat[Collaborative Learning]{\includegraphics[width = 3.5in]{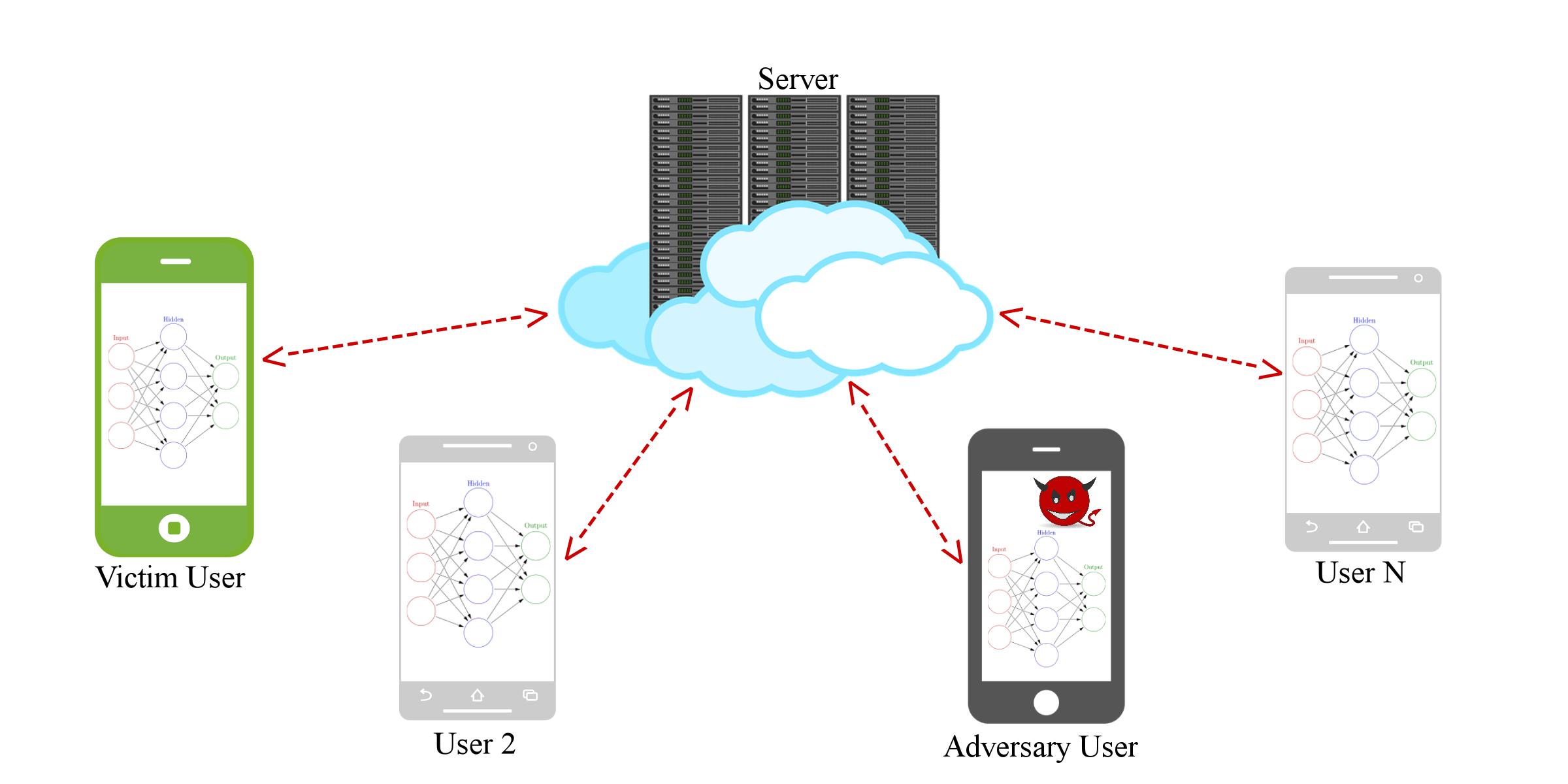}}
\end{tabular}
\caption{Two approaches for distributed deep learning. In (a), the red links show sharing of the data between the users and the server. Only the server can compromise the privacy of the data. In (b), the red links show sharing of the model parameters. In this case a malicious user employing a GAN can deceive any victim into releasing their private information.}
\label{fig:learningApproaches}
\end{figure*}

Deep Learning is a new branch of machine learning that makes use of neural networks, a concept which dates back to 1943 \cite{McCulloch1943}, to find solutions for a variety of complex tasks.
Neural networks were inspired by the way the human brain learns to show that distributed artificial neural networks could also learn nontrivial tasks, even though current architectures and learning procedures are far from brain-like behavior.

Algorithmic breakthroughs, the feasibility of collecting large amounts of data, and increasing computational power have contributed to the current popularity of neural networks, in particular with multiple (deep) hidden layers, that indeed have started to outperform previous state-of-the-art machine learning techniques \cite{BengioLearningDeepArchitectures, Goodfellow-et-al-2016-Book, Schmidhuber201585}. Unlike conventional machine learning approaches, deep learning needs no feature engineering of inputs \cite{lecun2015deep} since the model itself extracts relevant features on its own and defines which features are relevant for each problem \cite{lecun2015deep, Goodfellow-et-al-2016-Book}. 

Deep learning models perform extremely well with correlated data, which contributed to substantial improvements in computer vision \cite{lecun2010convolutional}, image processing, video processing, face recognition \cite{Taigman:2014:DCG:2679600.2680208}, speech recognition \cite{graves2013speech}, text-to-speech systems \cite{oord2016wavenet} and natural language processing \cite{leCunnNLP, DeepTextFCB, collobert2011natural}. 
Deep learning has also been used as a component in more complex systems that are able to play games  \cite{mnih2013playing, lai2015giraffe, googleGO, googleGoSedol} or diagnose and classify diseases \cite{deepmindHealthcare, cruz2013deep, fakoor2013using}. 

However, there are severe privacy implications associated with deep learning, as the trained model incorporates essential information about the training set. It is relatively straightforward to extract sensitive information from a model \cite{ateniese2015hacking, fredriksonMIAttack, fredrikson2014privacy}.

Consider the following cases depicted in Figure \ref{fig:learningApproaches}, in which $N$ users store local datasets of private information on their respective devices and would like to cooperate to build a common discriminative machine. We could build a classifier by uploading all datasets into a single location (e.g., the cloud), as depicted in Figure \ref{fig:learningApproaches} (a). A service operator trains the model on the combined datasets. This centralized approach is very effective since the model has access to all the data, but it's not privacy-preserving since the operator has direct access to sensitive information. We could also adopt a collaborative learning algorithm, as illustrated in Figure \ref{fig:learningApproaches} (b), where each participant trains a local model on his device and shares with the other users only a fraction of the parameters of the model. By collecting and exchanging these parameters, the service operator can create a trained model that is almost as accurate as a model built with a centralized approach. The decentralized approach is considered more privacy-friendly since datasets are not exposed directly. Also, it is shown experimentally to converge even in the case when only a small percentage of model parameters is shared and/or when parameters are truncated and/or obfuscated via differential privacy \cite{shokriPPDL}. But it needs several training passes through the data with users updating the parameters at each epoch.

The Deep Learning community has recently proposed Generative Adversarial Networks (GANs) \cite{NIPS2014_5423,Radford16,Salimans16}, which are still being intensively developed \cite{Bottou17, MeschederNG17, Bouchacourt16, Lamb16, Goodfellow15}. The goal of GANs is not to classify images into different categories, but to generate similar-looking samples to those in the training set (ideally with the same distribution). More importantly, GANs generate these samples without having access to the original samples. The GAN interacts only with the discriminative deep neural network to learn the distribution of the data.

In this paper, we devise a powerful attack against collaborative deep learning using GANs. The result of the attack is that any user acting as an insider can infer sensitive information from a victim's device. The attacker simply runs the collaborative learning algorithm and reconstructs sensitive information stored on the victim's device. The attacker is also able to influence the learning process and deceive the victim into releasing more detailed information. The attack works without compromising the service operator and even when model parameters are obfuscated via differential privacy. As depicted in Figure \ref{fig:learningApproaches}(a), the centralized server is the only player that compromises the privacy of the data. While in Figure \ref{fig:learningApproaches}(b), we show that any user can intentionally compromise any other user, making the distributed setting even more undesirable.

Our main contribution is to propose and implement a novel class of active inference attacks on deep neural networks in a collaborative setting. Our method is more effective than existing black-box or white-box information extraction mechanisms. 

Namely, our contributions are:
\begin{enumerate}
\item We devise a new attack on distributed deep learning based on GANs. GANs are typically used for implicit density estimation, and this, as far as we know, is the first application in which GANs are used maliciously.
\item Our attack is more generic and effective than current information extraction mechanisms. In particular, our approach can be employed against convolutional neural networks (CNN) which are notoriously difficult for model inversion attacks \cite{shokri2017membership}. 

\item We introduce the notion of deception in collaborative learning, where the adversary deceives a victim into releasing more accurate information on sensitive data. 
\item The attack we devise is also effective when parameters are obfuscated via differential privacy. We emphasize that it is not an attack against differential privacy but only on its proposed use in collaborative deep learning. In practice, we show that differentially private training as applied in \cite{shokriPPDL} and \cite{goodfellowDeepLearningwithDP} (example/record-level differential privacy) is ineffective in a collaborative learning setting under our notion of privacy.
\end{enumerate}

\section{Remarks}\label{remarks}
We devise a new attack that is more generic and effective than current information extraction mechanisms. It is based on Generative Adversarial Networks (GANs), which were proposed for implicit density estimation \cite{NIPS2014_5423}. The GAN, as detailed in Section \ref{background}, generates samples that appear to come from the training set, by pitting a generative deep neural network against a discriminative deep neural network. The generative learning is successful whenever the discriminative model cannot determine whether samples come from the GAN or the training set.  It is important to realize that both the discriminative and generative networks influence each other, because the discriminative algorithm tries to separate GAN-generated samples from real samples while the GAN tries to generate more realistic looking samples (ideally coming from the same distribution of the original data). The GAN never sees the actual training set, it only relies on the information stored in the discriminative model. The process is similar to the facial composite imaging used by police to identify suspects, where a composite artist {\em generates} a sketch from an eyewitness {\em discriminative} description of the face of the suspect. While the composite artist (GAN) has never seen the actual face, the final image is based on the feedback from the eyewitness. 

We use GANs in a new way, since they are used to extract information from honest victims in a collaborative deep learning framework. The GAN creates instances of a class that is supposed to be private. Our GAN-based method works only during the training phase in collaborative deep learning. 
Our attack is effective even against Convolutional Neural Networks which are notoriously difficult to invert  \cite{shokri2017membership}, or when parameters are obfuscated via differential privacy with granularity set at the record level (as proposed in \cite{shokriPPDL} and \cite{goodfellowDeepLearningwithDP}).
It works in a white-box access model where the attacker sees and uses internal parameters of the model. This in contrast to black-box access where the attacker sees only the output of the model for each particular input. It is not a limitation of our procedure because the purpose of collaborative learning is to share parameters, even if in a small percentage. 

Once the distributed learning process ends, a participant can always apply a model inversion or similar attack to the trained model. This is not surprising. 
What we show in this paper is that a malicious participant can see how the model evolves and influence other honest participants and force them into releasing relevant information about their private datasets. This ability to deceive honest users is unique to our attack. Furthermore, truncating or obfuscating shared parameters will not help since our attack is effective as long as the accuracy of the local models is high enough. 

\begin{figure}[!t]
\centering
\includegraphics[width=1.7in]{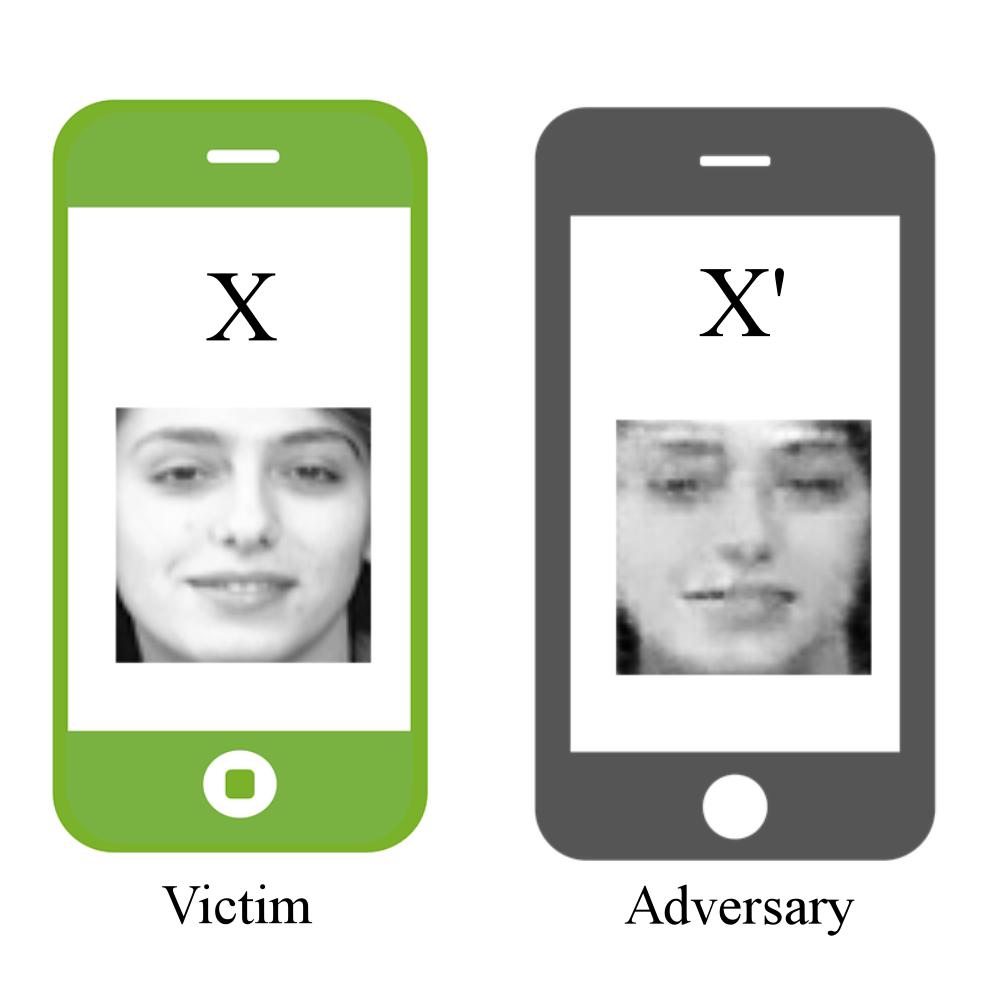}
\caption{Picture of Alice on the victim's phone, $X$, and its GAN reconstruction, $X'$. Note that $X'\neq X$, and $X'$ was not in the training set. But $X'$ is essentially indistinguishable from $X$.}
\label{victim_adversary_illustration}
\end{figure}

\begin{figure}[!t]
\centering
\includegraphics[width=1.7in]{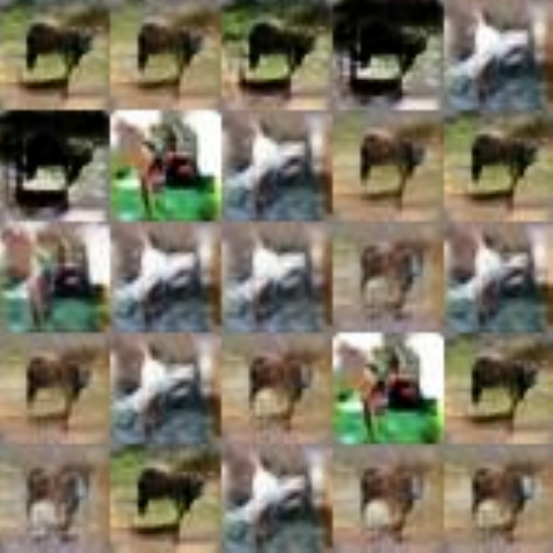}
\caption{GAN-generated samples for the `horse' class from the CIFAR-10 dataset}
\label{fig:HorseCifar10}
\end{figure}

{\em We emphasize however that our attack does not violate differential privacy (DP)}, which was defined to protect databases. The issue is that, in collaborative deep learning, DP is being applied to the parameters of the model and with granularity set at the record/example level. However, the noise added to learning parameters will ultimately have to be contained once the model becomes accurate. 
Our attack works whenever the model can accurately classify a class and will generate representatives of that class. The way DP is applied in \cite{shokriPPDL} and \cite{goodfellowDeepLearningwithDP} can at best protect against the recovery of specific elements associated with a label that was indeed used during the learning phase. 
The results of our attack may or may not be regarded as privacy violations.  Consider the following examples:

\begin{enumerate}
\item The victim's device contains standard medical records. The GAN will generate elements that look like generic medical records, i.e., items from the same distribution of those in the training set. The attacker may learn nothing of interest in this case, and there is no privacy violation. However, if the victim's device contains records of patients with cancer then the attacker may see inexistent patients, but all with cancer. Depending on the context, this may be considered a privacy violation.
\item The victim's device contains pornographic images. The GAN will generate similar scenes. While they may appear simulated, the information leaked to the adversary is significant. In other cases, our attack could be useful to law enforcement officials acting as adversaries. For instance, when the victim's device contains pedo-pornographic images or training material for terrorists.
\item The victim's device contains speech recordings. The GAN will generate babbling, with lots of fictitious word-like sounds (comparable to WaveNet \cite{oord2016wavenet} when the network is trained without the text sequence), thus there is no privacy violation. However, it may be possible to infer the language used (e.g., English or  Chinese) or whether the speaker is male or female, and this leaked information may constitute a privacy violation. 
\item The victim's device contains images of Alice. The GAN will generate faces that resemble Alice much like a composite artist generates a sketch of an eyewitness's memory of Alice. In our attack framework, the adversary will also collect all these drawings of Alice and falsely claim they are Eve's. This will force the local model within the victim's device to release more relevant and distinctive details about Alice's face, exacerbating the leakage. However, while many see this as a privacy violation, others may disagree since the adversary may not recover the exact face of Alice but only a reconstruction (see Figure \ref{victim_adversary_illustration}) .  On the other hand, if Alice wears glasses or has brown hair, then this information will be leaked and may constitute a privacy violation depending on the context. A further example is given in Figure \ref{fig:HorseCifar10}, where DCGAN was run on the CIFAR-10 dataset \cite{cifar10web} while targeting a class consisting of approximately 6,000 images containing various horses. Note that the class could be labeled `jj3h221f' and make no obvious reference to horses. The images produced by the GAN will tell the adversary that class `jj3h221f' does not contain cars or airplanes but animals (likely horses).
\end{enumerate}

Differential privacy in collaborative learning is meant to protect the recovery of specific elements used during training. Namely, an adversary cannot tell whether a certain $X$ was included in the training set (up to a certain threshold value). We circumvent this protection by generating an $X'$ which is indistinguishable from $X$. In Figure \ref{victim_adversary_illustration}, we show a real example of a face $X$ along with $X'$, the image generated by the GAN. Both images look similar even though $X'$ is not $X$. While this does not violate DP, it clearly leads to severe privacy violations in many cases. Our point is that example/record-level DP is inadequate in this context, much like secure encryption against a chosen-plaintext attack (CPA) is inadequate in an active adversarial environment. There is nothing wrong with DP per se (as there is nothing wrong with CPA-secure encryption); clearly DP provides information-theoretic protection but it's important to set its level of  granularity right. At record level, it is just not enough to protect sensitive information in collaborative learning against active adversaries. One can consider DP at different granularities (e.g., at user or device level) but this is not what is proposed in \cite{shokriPPDL}. Researchers can keep arguing about the proper use of DP or what DP is supposed to protect \cite{kifer2011no,frankDPLunchtime,freedomToTinker,frankDPcorrelated}, but ultimately, in the context of this work, one should ask: Would I use a system that let casual users recover images that are effectively indistinguishable from the ones in my picture folder? 

{\em The point is that collaborative learning for privacy is less desirable than the centralized learning approach it was supposed to improve upon: In centralized learning only the service provider can violate users' privacy, but in collaborative learning, any user may violate the privacy of other users in the system, without involving the service provider (see Figure \ref{fig:learningApproaches}).}

\section{Impact}
Google adopts a centralized approach and collects usage information from Android devices into a centralized database and runs machine learning algorithms on it. Google has recently introduced {\em Federated Learning} \cite{federatedDeepLearning, brendanRamageFederatedWeb} to enable mobile devices to collaboratively learn a shared prediction model while keeping all the training data local. Devices download the current model from a Google server and improve it by learning from local data. 

Federated learning appears to be the same as collaborative learning, and our attack should be equally effective. In the end, each device will download the trained model from the Google server, and the GAN will be able to operate successfully as long as the local model is learning. 

In federated learning, it is possible to protect individual model updates. Rather than using differential privacy as in \cite{shokriPPDL}, Google proposes to use a secure aggregation protocol. The updates from individual users' devices are securely aggregated by leveraging secure multiparty computation (MPC) to compute weighted averages of model parameters \cite{bonawitzpractical} so that the Google server can decrypt the result only if several users have participated. We believe that this mechanism, as described in their paper, is ineffective against our attack architecture since we simply rely on the fact that local models have successfully learned.  
Their security model considers only the case in which Google is the adversary that scrutinizes individual updates. Therefore, they don't consider the point we raise in this paper that casual users can attack other users. This makes federated learning potentially even more dangerous than the centralized one it is supposed to replace, at least in its current form. Indeed, our assessment is based on the description given in an announcement and two research papers. We have had no access to the actual implementation of the system yet, and products tend to improve significantly over time. 

Apple is said to apply differential privacy within a {\em crowdsourced learning} framework in future versions of iOS \cite{AppleDifferentialPrivacy}. While we do not know the details, we hope our paper serves as a warning on the risks of applying differential privacy improperly in collaborative deep learning. Our adversary does not have to work for the service provider, but he is a regular user targeting another user, e.g., a celebrity or a politician. 

\section{Related Work}
\label{relatedwork}
Deep Learning has proven to be successful in various areas of computer science. The capability to learn, process and produce relevant information from large quantities of data, makes deep learning a good option for the cyber security domain as well. However, new and unique attacks have emerged that pose a serious threat to the privacy of the information being processed. 

\subsection{Attacks on Machine Learning Models}

To the best of our knowledge, the first work that deals with extracting unexpected information from trained models is the one from Ateniese et al. \cite{ateniese2015hacking} (released in 2011 and on arXiv in 2013~\cite{Camenisch2012PrivacyOrientedC,ateniese2015hacking}). There, the authors devised a meta-classifier that is trained to {\em hack} into other machine learning classifiers to infer sensitive information or patterns from the training set. 
For instance, they were able to extract ethnicity or gender information from trained voice recognition systems.

The work was later extended by Fredrikson et al. \cite{fredriksonMIAttack,fredrikson2014privacy} where they proposed model inversion attacks on machine learning algorithms by exploiting confidence information revealed by the model. For instance, when applied to facial recognition systems, they show that it is possible to reconstruct images about a particular label known to the adversary. 

Recently, the work of Tram{\`e}r et al. \cite{tramer2016stealing} shows that \emph{stealing} machine learning models is possible when taking into consideration only the predictions provided by the model. Membership inference attacks were developed by Shokri et al. \cite{shokri2017membership}. Here, the adversary is given black-box access to the model and can infer whether a certain record was originally in the training set. 

McPherson et al. \cite{mcpherson2016defeating} use deep learning to infer and reveal the identity of subjects behind blurred images. In their work, Papernot et al. \cite{papernot2015limitations} show that an adversarially crafted input can be fed to deep learning models and make them prone to error, i.e., make the model misclassify the input therefore producing incorrect outputs. For example, a STOP sign on the road can be subtly modified to look the same to human eyes, but that is classified as another sign by a trained model. 
The work was extended in \cite{papernot2016transferability,grosse2016adversarial,xu2016automatically,laskovPracticalEvasion}.

\subsection{Privacy Preserving Machine Learning}

Defense mechanisms against powerful adversaries were devised by Shokri and Shmatikov \cite{shokriPPDL}. The authors introduce the concept of distributed deep learning as a way to protect the privacy of training data \cite{wainwright2012privacy}. In this model, multiple entities collaboratively train a model by sharing gradients of their individual models with each other through a parameter server.  Distributed learning is also considered in \cite{federatedDeepLearning, srinivasan2016distributed, mnih2016asynchronous, dean2012large, zhang2004solving, zinkevich2010parallelized}. Mohassel et al. \cite{mohasselsecureml}
 provide a solution for training neural networks while preserving the privacy of the participants. However, it deploys secure multiparty computation in the two-server model where clients outsource the computation to two untrusted but non-colluding servers. However, Shokri and Shmatikov \cite{shokriPPDL} are the \emph{first} to consider \emph{privacy-preserving measures with the purpose of finding practical alternatives to costly multi-party computation (MPC) techniques}.

Google developed techniques to train models on smartphones directly without transferring sensitive data to the company's data centers \cite{federatedDeepLearning, bonawitzpractical}. Microsoft developed CryptoNets \cite{CryptoNets:Microsoft} to perform deep learning on encrypted data and provide encrypted outputs to the users \cite{xie2014crypto}. Ohrimenko et al. \cite{ohrimenko2016oblivious} developed data-oblivious machine learning algorithms trained on trusted processors. Differential privacy plays an important role in deep learning as shown in \cite{goodfellowDeepLearningwithDP, shokriPPDL, kasiviswanathan2011can, song2013stochastic}.

\subsection{Differential Privacy}

Differential Privacy (DP) was introduced by Dwork \cite{dwork2006differential}. Its aim is to provide provable privacy guarantees for database records without significant query accuracy loss. Differential privacy for big data was considered by Dwork et al. \cite{dwork2014algorithmic}. Several works have adopted DP as an efficient defense mechanism  \cite{sarwate2013signal, diakonikolas, zhang2012functional, pathak2010multiparty, jain2015drop, bun2016concentrated, dwork2016concentrated, chaudhuri2011differentially, mcsherry2009differentially, bartheAdvancedDP, blocki2016differentially, narayan2015verifiable, eigner2014differentially}.

Collaborative deep learning proposed by Shokri and Shmatikov \cite{shokriPPDL} uses DP to obfuscate shared parameters while Abadi et al.  \cite{goodfellowDeepLearningwithDP} propose to apply DP to the parameters during training. DP was used in deep auto-encoders in \cite{phan2016differential}.

Covert channels, however, can be used to defeat DP-protected databases as shown in the work of Haeberlen et al. \cite{dpOnFire}.
In general, privacy cannot be guaranteed if auxiliary information (outside the DP model) is accessible to the 
adversary \cite{dwork2008difficulties}. At NDSS'16, it was shown by Liu et al. \cite{liu2016dependence} that DP at a certain granularity is not effective in real-life scenarios where data such as social data, mobile data, or medical records have strong correlations with each other. Note that it's a matter of setting DP granularity right and DP is not being violated at all. 

\subsection{Privacy-Preserving Collaborative Deep Learning}
\label{shokri}

A centralized approach to deep learning forces multiple participants to pool their datasets into a large central training set on which it is possible to train a model.  This poses serious privacy threats, as pointed out by Shokri and Shmatikov \cite{shokriPPDL}, and distrustful participants may not be willing to collaborate. 

Considering the security and privacy issues described above, Shokri and Shmatikov \cite{shokriPPDL} introduce a new collaborative learning approach, which allows participants to train their models, without explicitly sharing their training data. They exploit the fact that optimization algorithms, such as Stochastic Gradient Descent (SGD), can be parallelized and executed asynchronously. Their approach includes a selective parameter sharing process combined with local parameter updates during SGD. The participants share only a fraction of their local model gradients through a Parameter Server (PS). 
Each participant takes turns and uploads and downloads a percentage of the most recent gradients to avoid getting stuck into local minima.
This process only works if the participants agree in advance on a network architecture \cite{shokriPPDL}.

It is possible to blur the parameters shared with PS in various ways. Other than just uploading a small percentage of all the gradients, a participant can also select certain parameters that are above a threshold, within a certain range, or {\em noisy} in agreement with differential privacy procedures.

\section{Background} 
\label{background}

Supervised machine learning algorithms take labeled data and produce a classifier (or regressor) that it is able to accurately predict the label of new instances that has not seen before. Machine learning algorithms follow the inductive learning principle \cite{Vapnik98}, in which they go from a set examples to a general rule that works for any data coming from the same distribution as the training set. Given independent and identically distributed (i.i.d.) samples from $p(\mathbf{x},y)$, i.e., $\mathcal{D}=\{\mathbf{x}_i,y_i\}_{i=1}^n$, where $\mathbf{x}_i \in \mathbb{R}^d$ and $y_i\in\{1, 2, \ldots\}$, they solve the following optimization problem to find an accurate classifier:
\begin{equation}
\widehat{\theta}=\arg\min_{\theta\in \Theta} \sum_i L(f(\mathbf{x}_i; \theta) ,y_i) +\Omega(\theta),
\end{equation}
where $\hat{y}=f(\mathbf{x};\widehat{\theta})$ represents the learning machine, i.e., for any input $\mathbf{x}$ it provides an estimate for the class label $y$. $L(w, y)$ is a loss function that measures the error for misclassifying $y$ by $w$. And $\Omega(\theta)$ is a regularizer (independent of the training data) that avoids overfitting. Supervised learning algorithms like Support Vector Machines (SVMs) \cite{Schoelkopf02}, Random Forests \cite{Breiman01}, Gaussian Processes (GPs) \cite{Rasmussen06} and, of course, deep neural networks \cite{Goodfellow-et-al-2016-Book} can be depicted by this general framework.

Deep neural networks are becoming the weapon of choice when solving machine-learning problems for large databases with high-dimensional strongly correlated inputs because they are able to provide significant accuracy gains. Their improvements are based on additionally learning the features that go into the classifier. Before deep learning, in problems that dealt with high-dimensional strongly correlated inputs (e.g., images or voice), humanly engineered features, which were built to reduce dimensionality and correlation, were fed to a classifier of choice. The deep neural network revolution has shown that the features should not be humanly engineered but \emph{learned from the data}, because the hand-coded features were missing out relevant information to produce optimal results for the available data. The deep neural network learns the useful features that make sense for each problem, instead of relying on best guesses. The deep neural network structures are designed to exploit the correlation in the input to learn the features that are ideal for optimal classification. The deep structure is needed to extract those features in several stages, moving from local features in the lower layers to global features at the higher layers, before providing an accurate prediction on the top layer. These results have become self-evident when datasets have grown in size and richness. 

The learning machine $f(\mathbf{x};\theta)$ summarizes the training database in the estimated parameters $\widehat{\theta}$. From the learning machine and its estimated parameters, relevant features of the training database, if not complete training examples, can be recovered. So an adversary that wants to learn features from the original training data can do so if it has access to the learning machine. For example, SVMs store prototypical examples from each class in $\widehat{\theta}$ and GPs store all the training points, so there is no challenge there for an adversary to learn prototypical examples for each class in those classifiers. For deep neural networks, the relation between $\widehat{\theta}$ and the training points in $\mathcal{D}$ is more subtle, so researchers have tried to show that privacy is a possibility in these networks \cite{shokriPPDL}. But the model inversion attack \cite{fredriksonMIAttack,fredrikson2014privacy} has proven that we can recover inputs (e.g., images) that look similar to those in the training set, leaking information to the adversary about how each class looks like. And as deep neural networks are trained with unprocessed inputs, these attacks recover prototypical examples of the original inputs. 

It is important to emphasize that this is an intrinsic property of any machine-learning algorithm. If the algorithm has learned and it is providing accurate classification, then an adversary with access to the model can obtain information from the classes. If the adversary has access to the model, it can recover prototypical examples from each class. If sensitive or private information is needed for the classifier to perform optimally, the learning machine can potentially leak that information to the adversary. We cannot have it both ways, either the learning machine learns successfully, or data is kept private. 

\subsection{Limitations of the Model Inversion Attack}

The model inversion attack works in a simple way \cite{fredriksonMIAttack,fredrikson2014privacy}: Once the network has been trained, we can follow the gradient used to adjust the weights of the network and obtain a reverse-engineered example for all represented classes in the network. For those classes that we did not have prior information, we would still be able to recover prototypical examples. This attack shows that any accurate deep learning machine, no matter how it has been trained, can leak information about the different classes that it can distinguish.

Moreover, the model inversion attack may recover only prototypical examples that have little resemblance to the actual data that defined that class. This is due to the rich structure of deep learning machines, in which broad areas of the input space are classified with high accuracy but something else is left out \cite{42503,Shlens15}. If this is the case, the adversary might think he has recovered sensitive information for that class when he is just getting meaningless information. For example, we refer the reader to Figure 5 from \cite{42503}, where six training images for a school bus, bird, a temple, soap dispenser, a mantis and a dog have been slightly tweaked to be classified as an ostrich (Struthio camelus), while they still look like the original image. In \cite{Shlens15}, the authors show in Figure 5 a procedure similar to the model inversion attack. A randomly generated image, plus gradient information from the deep belief network, produces a random looking image that is classified as an airplane. The structure of deep neural networks is so large and flexible that it can be fooled into giving an accurate label even though the image to a human looks nothing like it. 

Thus any model inversion attack can obtain private information from a trained deep neural network, but it can land in an unrepresented part of the input space that looks nothing like the true inputs defined for each class. Extensive research in the ML community has shown that GAN generated samples are quite similar to the training data, thus the results coming from our attack reveal more sensitive information about the training data compared to the average samples or aggregated information one would expect from a model inversion type of attack.

\subsection{Generative Adversarial Networks}

One way to address the problem highlighted in \cite{42503,Shlens15} is generating more training images so to cover a larger portion of the space. This can be accomplished through Generative Adversarial Networks (GANs) \cite{NIPS2014_5423}.

The GAN procedure pits a discriminative deep learning network against a generative deep learning network. In the original paper \cite{NIPS2014_5423}, the discriminative network is trained to distinguish between images from an original database and those generated by the GAN. The generative network is first initialized with random noise, and at each iteration, it is trained to mimic the images in the training set of the discriminative network. The optimization problem solved by the GAN procedure can be summarized as
\begin{equation}\label{GAN}
\min_{\theta_G}\max_{\theta_D} \sum_{i=1}^{n_+} \log f(\mathbf{x}_i;\theta_D)+\sum_{j=1}^{n_-} \log(1-f(g(\mathbf{z}_j;\theta_G);\theta_D))
\end{equation}
where $\mathbf{x}_i$ are images from the original data and $\mathbf{z}_j$ are randomly generated images (e.g., each pixel distributed between 0 and 255 uniformly). Let $f(\mathbf{x};\theta_D)$ be a discriminative deep neural network that, given an image, produces a class label and let $\theta_D$ denote its parameters. Let $g(\mathbf{z};\theta_G)$ be a generative deep neural network, which given a random input produces an image. 

The training procedure works as follows. First, we compute the gradient on $\theta_D$ to maximize the performance of the discriminative deep neural network. Hence $f(\mathbf{x};\theta_D)$ is able to distinguish between samples from the original data, i.e., $\mathbf{x}_i$, and samples generated from the generative structure, i.e., $\mathbf{x}_j^{\text{fake}}=g(\mathbf{z}_j;\theta_G)$. Second, we compute the gradients on $\theta_G$, so the samples generated from $\mathbf{x}_j^{\text{fake}}=g(\mathbf{z}_j;\theta_G)$ look like a perfect replica of the original data\footnote{The generated data looks like the original data, but they are not copies of them.}. 

The procedure ends when the discriminative network is unable to distinguish between samples from the original database and the samples generated by the generative network. The authors of the paper \cite{NIPS2014_5423} prove the following theorem:
\vspace*{.2cm}
\begin{theorem}
The global minimum of the virtual training criterion in \eqref{GAN} is achieved if and only if $p(\mathbf{x}) = p(g(\mathbf{z};\theta_G))$.
\end{theorem}
\vspace*{.2cm}

The theorem shows that the adversarial game ends when the GAN is generating images that appear to come from the original dataset. 

In \cite{Goodfellow15}, the author shows that in the infinite sample limit the generative network would draw samples from the original training distribution. But it also recognizes that the GAN procedure will not converge. In a recent paper \cite{Salimans16}, the authors have significantly improved the training of the GAN including new features to improve convergence to the density model.

\section{Threat Model}
\label{threat}

Our threat model follows \cite{shokriPPDL}, but  relies on an active insider.
 
The adversary pretends to be an honest participant in the collaborative deep learning protocol but tries to extract information about a class of data he does not own. 
The adversary will also surreptitiously \emph{influence} the learning process to deceive a victim into releasing further details about the targeted class. This \emph{adversarial influence} is what makes our attack more effective than, for instance, just applying model inversion attacks \cite{fredriksonMIAttack} against the final trained model. Furthermore, our attack works for more general learning models (those for which a GAN can be implemented), including those on which model inversion attack is notoriously ineffective (e.g., convolutional neural networks).

\begin{figure*}
  \includegraphics[width=\linewidth]{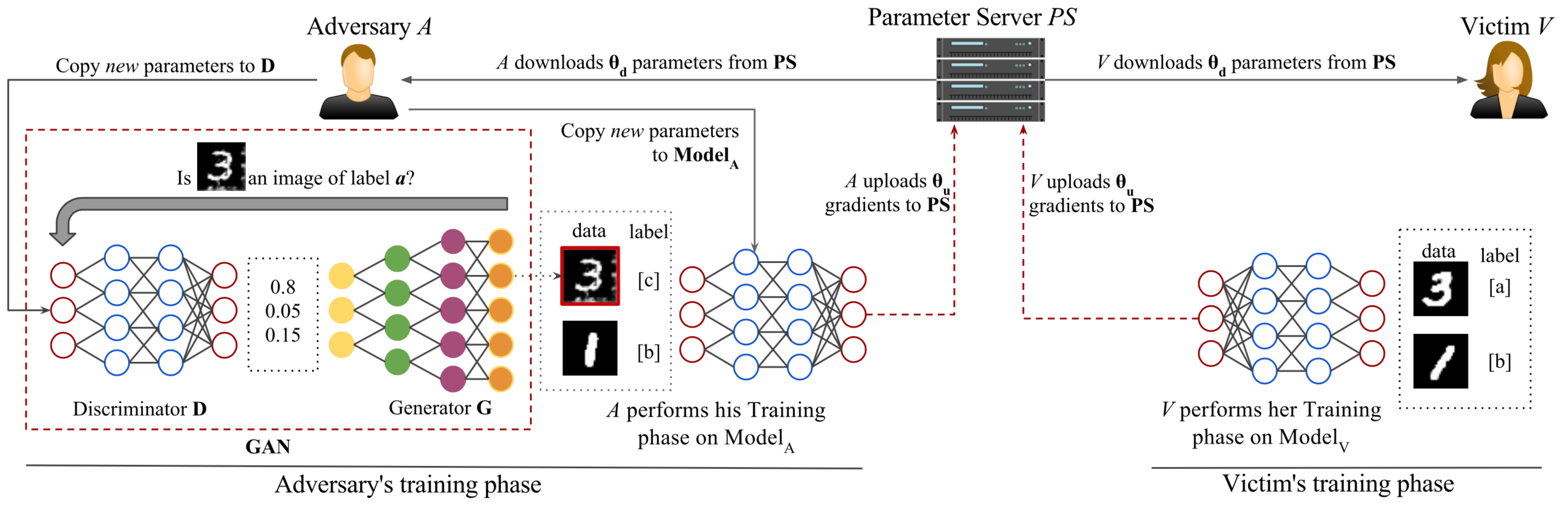}
  \caption{GAN Attack on collaborative deep learning. The victim on the right trains the model with images of 3s (class $a$) and images of 1s (class $b$). The adversary only has images of class $b$ (1s) and uses its label $c$ and a GAN to fool the victim into releasing information about class $a$. The attack can be easily generalized to several classes and users. The adversary does not even need to start with any true samples.}
  \label{fig:ganscenario}
\end{figure*}

Specifically, we consider the following scenario:
\begin{itemize} 
 \item The adversary works as an insider within the privacy-preserving collaborative deep learning protocol.
 \item The objective of the adversary is to infer meaningful information about a label that he does not own.
 \item The adversary does not compromise the central parameter server (PS) that collects and distributes parameters to the participants. That is, the parameter server, or the service provider in our example, is not under the control of the adversary. In our real-world example, the adversary is a full-fledged insider and does not have to work for the service provider. 
 \item The adversary is {\em active} since he directly manipulates values and builds a GAN locally. At the same time, he follows the protocol specification as viewed by his victims. In particular, the adversary takes turns, follows the parameter selection procedures, uploads and downloads the correct amount of gradients as agreed in advance, and obfuscates the uploaded parameters as required by the collaborative learning process.
 \item As in \cite{shokriPPDL}, it is assumed that {\it all} participants agree in advance on a \emph{common learning objective}. This implies that the adversary has knowledge of the model structure and, in particular, of the data labels of other participants.
 \item Unlike static adversaries as in model inversion \cite{fredriksonMIAttack}, our adversary is allowed to be {\em adaptive} and work in real time while the learning is in progress. The adversary will be able to influence other participants by sharing specially-crafted gradients and trick participants into leaking more information on their local data. This is possible because the distributed learning procedure needs to run for several rounds before it is successful. 
\end{itemize}

\section{Proposed Attack}
\label{attack}
The adversary $A$ participates in the collaborative deep learning protocol. All participants agree in advance on a common learning objective \cite{shokriPPDL} which means that they agree on the type of neural network architecture and on the labels on which the training would take place. 

Let $V$ be another participant (the victim) that declares labels $[a, b]$. The adversary $A$ declares labels $[b, c]$. Thus, while $b$ is in common, $A$ has no information about the class $a$. The goal of the adversary is to infer as much useful information as possible about elements in $a$. 

Our insider employs a GAN to generate instances that look like the samples from class $a$ of the victim. The insider injects these fake samples from $a$, as class $c$ into the distributed learning procedure. In this way, the victim needs to work harder to distinguish between classes $a$ and $c$ and hence will reveal more information about class $a$ than initially intended. Thus, the insider mimics samples from $a$ and uses the victim to improve his knowledge about a class he ignored before training. GANs were initially devised for density estimation, so we could learn the distribution of the data from the output of a classifier without seeing the data directly. In this case, we use this property to deceive the victim into providing more information about a class that is unknown to the insider. 

For simplicity, we consider first two players (the adversary and the victim) and then extend our attack strategy to account for multiple users.  Each player can declare any number of labels, and there is no need for the classes to overlap. 

\begin{enumerate}
\item Assume two participants $A$ and $V$. Establish and agree on the common learning structure and goal.
\item $V$ declares labels $[a, b]$ and $A$ labels $[b, c]$.
\item  Run the collaborative deep learning protocol for several epochs and stop only when the model at the parameter server (PS) and both local models have reached an accuracy that is higher than a certain threshold.
\item First, the Victim trains the network: 
\begin{enumerate}
\item $V$ downloads a percentage of parameters from PS and updates his local model.
\item $V$'s local model is trained on $[a, b]$.
\item $V$ uploads a selection of the parameters of his local model to PS.
\end{enumerate}
\item Second, the Adversary trains the network:
\begin{enumerate}
\item $A$ downloads a percentage of parameters from the PS and update his local model. 
\item {\it $A$ trains his local generative adversarial network (unknown to the victim) to mimic class $a$ from the victim. }
\item {\it $A$  generates samples from the GAN and labels them as class $c$.}
\item $A$'s local model is trained on $[b, c]$.
\item $A$ uploads a selection of the parameters of his local model to PS.   
\end{enumerate}
\item Iterate between 4) and 5) until convergence. 
\end{enumerate}

The steps highlighted in 5b) and 5c) above represent the extra work the adversary perform to learn as much as possible elements of the targeted label $a$.  The procedure is depicted in Figure \ref{fig:ganscenario}. The generalization of the attack to multiple users is reported in Algorithm \ref{alg:pseudo}.

\begin{algorithm}
\caption{Collaborative Training under GAN attack}
\label{alg:pseudo}
 \begin{algorithmic}[1]
 \renewcommand{\algorithmicrequire}{\textbf{Pre-Training Phase:}}
 \renewcommand{\algorithmicensure}{\textbf{Training Phase}}
 \REQUIRE Participants agree in advance on the following, as pointed out also by \cite{shokriPPDL}:
   \begin{enumerate}
     \item common learning architecture, (model, labels etc.) \COMMENT{For ex. $V$ declares labels $[a, b]$ and $A$ labels $[b, c]$}
    \item learning rate, (lr)
    \item parameter upload fraction (percentage), $(\theta_u)$
    \item parameter download fraction, $(\theta_d)$
    \item threshold for gradient selection, ($\tau$)
    \item bound of shared gradients, ($\gamma$)
    \item training procedure, (sequential, asynchronous)
    \item parameter upload criteria \COMMENT{cf. \cite{shokriPPDL}}
   \end{enumerate}
 \ENSURE 
  \FOR {$epoch = 1$ to $nrEpochs$}
  \STATE Enable user $X$ for training
  \STATE User $x$ downloads $\theta_d$ parameters from PS
  \STATE Replace respective local parameters on user $x$ local model with newly downloaded ones
  \IF {($user\_type ==  ADVERSARY$)}
  \STATE Create a replica of local $freshly updated$ model as $D$ (discriminator)
  \STATE Run Generator $G$ on $D$ targeting class $a$ (unknown to the adversary)
  \STATE Update $G$ based on the answer from $D$
  \STATE Get n-samples of class $a$ generated by $G$
  \STATE Assign label $c$ (fake label) to generated samples of class $a$
  \STATE Merge the generated data with the local dataset of the adversary
  \ENDIF
  \STATE Run SGD on local dataset and update the local model
  \STATE Compute the gradient vector $(new Parameters - old Parameters)$
  \STATE Upload $\theta_u$ parameters to PS
  \ENDFOR
 \RETURN Collaboratively Trained Model \COMMENT{At the end of training, the adversary will have prototypical examples of members of class $a$ known only to the victim}
 \end{algorithmic} 
 \end{algorithm}

The GAN attack works as long as $A$'s local model improves its accuracy over time. 
Another important point is that the GAN attack works even when differential privacy or other obfuscation techniques are employed.  It is not an attack on differential privacy but on its proposed use in collaborative deep learning. Though there might be a degradation in the quality of results obtained, our experiments show that as long as the model is learning, the GAN can improve and learn, too. Of course, there may always exist a setup where the attack may be thwarted. This may be achieved by setting stronger privacy guarantees, releasing fewer parameters, or establishing tighter thresholds. However, as also shown by the results in \cite{shokriPPDL}, such measures lead to models that are unable to learn or that perform worse than models trained on centralized data.  In the end, the attack is effective even when differential privacy is deployed, because the success of the generative-discriminative synergistic learning relies only on the accuracy of the discriminative model and not on its actual gradient values.

\section{Experimental Setup}
\label{experimentalsetup}

The authors of \cite{shokriPPDL} provided us with their source code that implements a complete distributed collaborative learning system. Our attacks were run using their implementation of differential privacy.

\subsection{Datasets}

We conducted our experiments on two well-known datasets, namely MNIST \cite{lecunMnistWeb} and AT\&T dataset of faces \cite{samaria1994parameterisation} (a.k.a. Olivetti dataset of faces).
\subsubsection{MNIST Dataset of Images} MNIST is the benchmark dataset of choice in several deep learning applications. It consists of handwritten grayscale images of digits ranging from 0 to 9. Each image is of $32\times32$ pixels and centered. The dataset consists of 60,000 training data records and 10,000 records serving as test data.
\subsubsection{AT\&T Dataset of Faces (Olivetti dataset)}
AT\&T dataset, previously used also in the work of \cite{fredriksonMIAttack}, consists of grayscale images of faces of several persons taken in different positions. The version used in our experiments consists of 400 images of 64$\times$64 pixels.\footnote{http://www.cs.nyu.edu/\~roweis/data.html}  The dataset contains images of 40 different persons, namely 10 images per person. \\ \\

For these experiments, we did not conduct any pre-processing of the data. The only processing performed on the data was scaling every image to the $[-1, +1]$ range, similar to \cite{Radford16}. This was done to adopt the state-of-the-art generator model of \cite{Radford16}, which has a hyperbolic tangent $tanh$ activation function in its last layer, thus outputting results in the $[-1, +1]$ range as well.

\subsection{Framework}

We build our experiments on the Torch7 scientific computing framework.\footnote{http://torch.ch/} Torch is one of the most widely used deep learning frameworks. It provides fast and efficient construction of deep learning models thanks to LuaJIT\footnote{http://luajit.org}, a scripting language which is based on Lua\footnote{https://www.lua.org}. 

\subsection{System Architecture}
\label{systarchitecture}

We used a Convolutional Neural Network (CNN) based architecture during our experiments on MNIST and AT\&T. The layers of the networks are sequentially attached to one another based on the $nn.Sequential()$ container so that layers are in a feed-forward fully connected manner.\footnote{https://github.com/torch/nn/blob/master/doc/containers.md\#nn.Sequential}

In the case of MNIST (Figure \ref{cnn_architecture_mnist}), the model consists of two convolution layers, $nn.SpatialConvolutionMM()$, where the $tanh$ function is applied to the output of each layer before it is forwarded to the max pooling layers, $nn.SpatialMaxPooling()$. The first convolutional layer has a convolution kernel of size 5$\times$5 and it takes one input plane and it produces 32 output planes. Whereas the second convolutional layer takes 32 input planes and produces 64 output planes and it has a convolution kernel of size 5$\times$5. After the last max pooling layer, the data gets reshaped on a tensor of size 256, on which a linear transformation is applied which takes as input the tensor of size 256 and outputs a tensor of size 200. Then a $tanh$ activation function is applied to the output, which is then followed by another linear transformation which takes as input the tensor of size 200 and outputs a tensor of size 11. We modify the output layer from 10 to 11, where the 11th output is where the adversary trains with the results generated by $G$. As in Goodfellow et. al \cite{NIPS2014_5423}, the 11th class is the class where the `fake' images are placed. Further details are provided on Section \ref{experiments}. The last layer of the models is a LogSoftMax layer, $nn.LogSoftMax()$.

Images in the AT\&T dataset of faces are larger (64$\times$64). Therefore, we built a convolutional neural network (Figure \ref{cnn_architecture_att}) consisting of three convolution layers and three max pooling layers, followed by the fully connected layers in the end. As in the MNIST architecture, $tanh$ is used as an activation function. This model has an output layer of size 41, namely 40 for the real data of the persons and 1 as the class where the adversary puts the reconstructions for his class of interest. Since faces are harder to reconstruct than numbers, we implemented Algorithm \ref{alg:pseudo} differently. For this case, the generator $G$ queries the discriminator $D$ more times per epoch (size of adversary's training data divided by batch size) to improve faster. 

The $Generator (G)$ architecture used in MNIST-related experiments, Figure \ref{g_architecture}, consisted of 4 convolution layers corresponding to $nn.SpatialFullConvolution()$ from the torch `nn' library. 
Batch normalization, \emph{nn.SpatialBatchNormalization()}, is applied to the output of all layers except the last one. The activation function is the rectified linear unit function, $nn.ReLU()$.
The last layer of the model is a hyperbolic tangent function, $tanh$, to set the output of $G$ to the [-1, +1] range. Since AT\&T images are larger (64x64), $G$ has an additional (5th) convolution layer.
The number of convolution layers needed were computed automatically using the techniques from \cite{perarnau2016invertible}. $G$ takes as input a 100-dimensional uniform distribution \cite{Radford16, chintalaTorchCode}, and converts it to a 32x32 image for MNIST or a 64x64 image for AT\&T. 
As in \cite{chintalaTorchCode}, we initialized the weights of the generator with 0 mean and 0.02 standard deviation. While \cite{Radford16} applies this initialization function to both $D$ and $G$, we do it \emph{only} to $G$ since $D$ is the model that is shared among all participants. 

Both architectures described above are represented in Figure \ref{g_architecture} and \ref{fig:dcgan_faces_generator} as printed out by Torch7.

We refer the reader to Appendix \ref{sysarch} for further details on the architectures provided by Torch7.

\subsection{Hyperparameter Setup}

For the MNIST-related experiments, we set the learning rate for both the collaboratively trained model and the discriminator model to $1e-3$, learning rate decay of $1e-7$, momentum $0$ and batch size of $64$.

For the AT\&T-related experiments, we set the learning rate to $0.02$ and a batch size of $32$. Whereas, for the AT\&T experiments concerning the multi-participant scenario, we used a batch size of $1$. We kept the rest of the hyperparameters similar to the MNIST case. A learning rate of $0.02$ worked better as it allowed more stochasticity in the process, thus allowing the model to converge faster.

The authors of DCGAN \cite{Radford16} use the $Adam$ optimizer with a learning rate of $0.0002$ and a momentum term $\beta_1$ of 0.5 as provided in the torch implementation of DCGAN \cite{chintalaTorchCode}. We modified the process to use stochastic gradient descent (SGD) and, for this configuration, a learning rate of $0.02$ for the generator worked better.

\section{Experiments}
\label{experiments}

We now evaluate how well our GAN procedure can recover records from other participants. 
We focus our experiments on MNIST and AT\&T datasets that contain images. In principle, however, our adversarial strategy can be extended to other types of data, such as music, audio, medical records, etc.
We first compare our GAN attack against model inversion in a traditional setting. As mentioned before, model inversion has several limitations and may not be effective against certain types of neural networks. While this may be clear from a theoretical perspective, we also provide experimental evidence for this claim in the first experiment. 

In the second set of experiments, we show how the GAN attack also works in the distributed setting in which the adversary is oblivious to the content of some, or all, labels, see Figure \ref{ppdl_architecture}. 

In the third set of experiments, we show that adding noise to the parameters of the deep neural network before they are uploaded to the parameter server does not protect against our GAN attack. In general, deploying record-level differential privacy to obfuscate the model parameters is ineffective against our attack. The efficacy of the GAN is only limited by the accuracy of the discriminator.

\subsection{MI Attack vs. GAN Attack}

In this first example, we compare the model inversion (MI) and the GAN attacks, and we provide them with all the data. The adversary has access to the fully trained models. 

\begin{figure}
\centering
\resizebox{2.5in}{!}{%
\begin{tabular}{c|c|c}
Actual Image & MIA & DCGAN \\ \hline
\includegraphics[width = 0.3in]{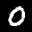} & \includegraphics[width = 0.3in]{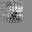} & \includegraphics[width = 0.3in]{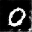}          \\
\includegraphics[width = 0.3in]{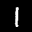} & \includegraphics[width = 0.3in]{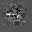} & \includegraphics[width = 0.3in]{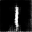}          \\
\includegraphics[width = 0.3in]{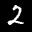} & \includegraphics[width = 0.3in]{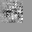} & \includegraphics[width = 0.3in]{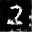}          \\
\includegraphics[width = 0.3in]{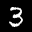} & \includegraphics[width = 0.3in]{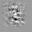} &  \includegraphics[width = 0.3in]{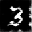}         \\
\includegraphics[width = 0.3in]{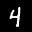} & \includegraphics[width = 0.3in]{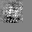} &  \includegraphics[width = 0.3in]{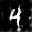}         \\
\includegraphics[width = 0.3in]{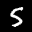} & \includegraphics[width = 0.3in]{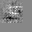} & \includegraphics[width = 0.3in]{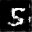}         \\         
\includegraphics[width = 0.3in]{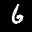} & \includegraphics[width = 0.3in]{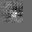} & \includegraphics[width = 0.3in]{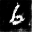}        \\
\includegraphics[width = 0.3in]{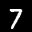} & \includegraphics[width = 0.3in]{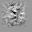} &  \includegraphics[width = 0.3in]{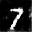}        \\
\includegraphics[width = 0.3in]{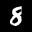} & \includegraphics[width = 0.3in]{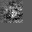} & \includegraphics[width = 0.3in]{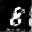}           \\
\includegraphics[width = 0.3in]{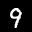} & \includegraphics[width = 0.3in]{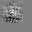} & \includegraphics[width = 0.3in]{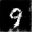}
\end{tabular}%
}
\caption{Results obtained when running model inversion attack (MIA) and a generative adversarial network (DCGAN) on CNN trained on the MNIST dataset. MIA fails to produce clear results, while DCGAN is successful.}
\label{miavsgan}
\end{figure}

For the MI attack, we train a convolutional neural network on all 60,000 training examples of the MNIST dataset. We apply the model inversion attack in \cite{fredriksonMIAttack}, once the deep neural network is trained. However, instead of approximating the derivatives as in \cite{fredriksonMIAttack}, we collected the exact gradients computed by the model on the input given and the label (class) of interest.  The results are shown in Figure \ref{miavsgan}. MI works well for MLP networks but clearly fails with CNNs. This is consistent with the work \cite{shokri2017membership} where the authors attained similar results.  It appears that MI is not effective when dealing with more complicated learning structures. While relevant information is in the network, the gradients might take us to an area of the input space that is not representative of the data that we are trying to recover. 

For the GAN approach, we adopt the DCGAN architecture in \cite{Radford16}, and its torch implementation from \cite{chintalaTorchCode}. The model consists of the discriminator (D) in combination with the DCGAN generator (G). We made the generator model compatible with MNIST-type of images and used methods proposed in \cite{perarnau2016invertible} so that our code could automatically calculate the number of convolution layers needed. We refer the reader to Section \ref{systarchitecture} for further details on the architectures. We ran the experiments 10-times (once per each class present in the MNIST dataset), and we let the models train until the accuracy reached by D was above $97\%$. We show the results in Figure \ref{miavsgan}. 

Note a significant difference: In the GAN attack, the generative model is trained together with the discriminative model, while in MI, the discriminative model is only accessed at the end of the training phase. However, this type of \emph{real-time access} to the model is what makes our attack applicable to collaborative deep learning. 

\subsection{GAN Attack on Collaborative Learning without Differential Privacy}
\label{ganattack}

\begin{figure*}
\centering
\setlength{\tabcolsep}{2pt}
\begin{tabular*}{\textwidth}{c c c}
\setlength{\tabcolsep}{1pt}
\subfloat[$\theta_u = 1, \theta_d = 1$]{
\begin{tabular}{lllll}

\includegraphics[width = 0.4in]{original_for_sample_5_400_100.png}  & \includegraphics[width = 0.4in]{original_for_sample_5_390_200.png} & \includegraphics[width = 0.4in]{original_for_sample_5_500_300.png} & \includegraphics[width = 0.4in]{original_for_sample_5_480_400.png} & \includegraphics[width = 0.4in]{original_for_sample_5_400_500.png} \\ 
\includegraphics[width = 0.4in]{sample_5_epoch_400_taskid_100.png}  & \includegraphics[width = 0.4in]{sample_5_epoch_390_taskid_200.png} & \includegraphics[width = 0.4in]{sample_5_epoch_500_taskid_300.png} & \includegraphics[width = 0.4in]{sample_5_epoch_480_taskid_400.png} & \includegraphics[width = 0.4in]{sample_5_epoch_400_taskid_500.png}  \\
\end{tabular}
}
    &
 \setlength{\tabcolsep}{1pt}
\subfloat[$\theta_u = 0.1, \theta_d = 1$]{
\begin{tabular}{lllll}

\includegraphics[width = 0.4in]{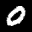}  & \includegraphics[width = 0.4in]{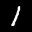} & \includegraphics[width = 0.4in]{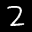} & \includegraphics[width = 0.4in]{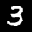} & \includegraphics[width = 0.4in]{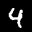} \\ 
\includegraphics[width = 0.4in]{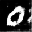}  & \includegraphics[width = 0.4in]{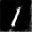} & \includegraphics[width = 0.4in]{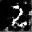} & \includegraphics[width = 0.4in]{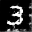} & \includegraphics[width = 0.4in]{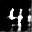} \\ 
\end{tabular}
}
    &
 \setlength{\tabcolsep}{1pt}
\subfloat[$\theta_u = 0.1, \theta_d = 0.1$]{
\begin{tabular}{lllll}

\includegraphics[width = 0.4in]{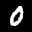}  & \includegraphics[width = 0.4in]{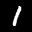} & \includegraphics[width = 0.4in]{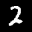} & \includegraphics[width = 0.4in]{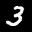} & \includegraphics[width = 0.4in]{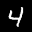}\\ 
\includegraphics[width = 0.4in]{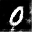}  & \includegraphics[width = 0.4in]{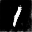} & \includegraphics[width = 0.4in]{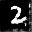} & \includegraphics[width = 0.4in]{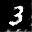} & \includegraphics[width = 0.4in]{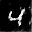} \\ 
\end{tabular}
}
\end{tabular*}
\caption{Results for the GAN attack on a two-user scenario. Bottom row, samples generated by the GAN. Top row, samples from the training set closest to the ones generated by the GAN. (a) 100\% parameters upload and download. (b) 100\% download and 10\% upload. (c) 10\% upload and download.}
\label{fig:ganresults}
\end{figure*}

\begin{figure*}
\centering
\includegraphics[width=\linewidth]{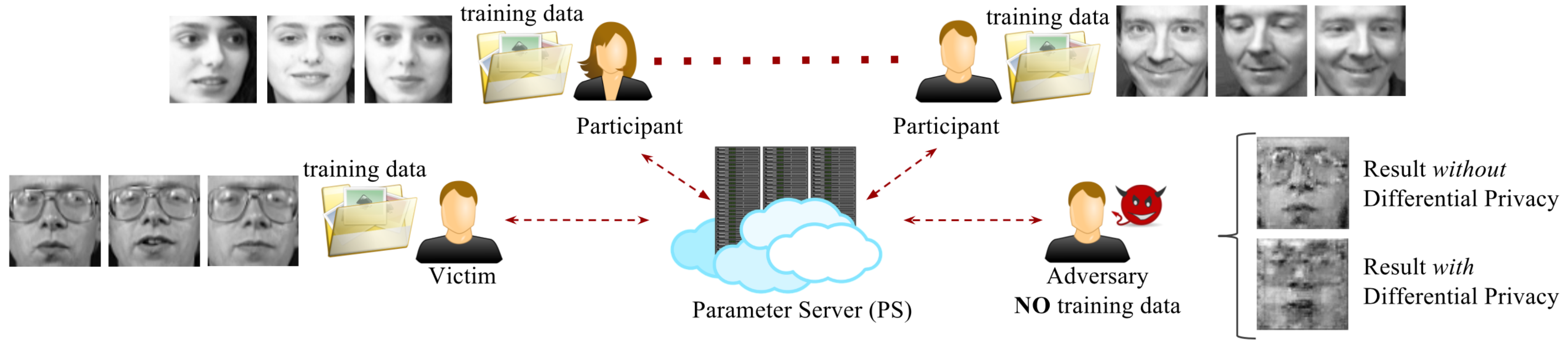}
\caption{Collaborative deep learning with 41 participants. All 40 honest users train their respective models on distinct faces. The adversary has no local data. The GAN on the adversary's device is able to reconstruct the face stored on the victim's device (even when DP is enabled).}
\label{ppdl_architecture}
\end{figure*}

Now we set the GAN attack in a collaborative environment like the one proposed in \cite{shokriPPDL}. We use the model described in Section \ref{attack} and depicted in Figure \ref{fig:ganscenario}. 
\subsubsection{Experiments on MNIST}
Instead of using two labels per user, we use five labels for the first user and six labels for the second user. The first user has access to images of 0 to 4 (with label 1 to 5) and the second user, the adversary, has access to images of 5 to 9 (label 6 to 10). The adversary uses its sixth class to extract information on one of the labels of the first user.

The results are shown in Figure \ref{fig:ganresults}. 
For every retrieved image (bottom row), we placed above it an actual training image from the first user (we show the image that is closest in L1-norm). We have repeated the experiment with three different parameter settings. In (a), the users upload and download the entire model. In (b), the users download the full model, but only upload 10\% of the parameters in each epoch. Finally, in (c), the upload and download is only 10\%.

\begin{figure}
\centering
\captionsetup[subfigure]{labelformat=empty, justification=centering}
\begin{tabular}{c|ccc}
\centering
\subfloat[Original]{\includegraphics[width = 0.6in]{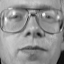}} &
\subfloat[$\theta_u = 1 \newline \theta_d = 1$]{\includegraphics[width = 0.6in]{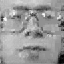}} &
\subfloat[$\theta_u = 0.1 \newline \theta_d = 1$]{\includegraphics[width = 0.6in]{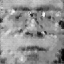}} &
\subfloat[$\theta_u = 0.1 \newline \theta_d = 0.1$]{\includegraphics[width = 0.6in]{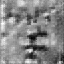}} 
\end{tabular}

\caption{Experimental results on the AT\&T Dataset with no DP. Unlike MNIST, images are noisier because this particular dataset is small and the accuracy of the model is significantly affected when upload rates are small.}
\label{fig:OlivettiNoDP}
\end{figure}

\begin{figure*}
\centering
\captionsetup[subfigure]{labelformat=empty, justification=centering}
\begin{tabular}{cccccccccccc}
\centering
\subfloat[epoch 5]{\includegraphics[width = 0.4in]{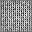}} &
\subfloat[epoch 20]{\includegraphics[width = 0.4in]{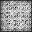}} &
\subfloat[epoch 35]{\includegraphics[width = 0.4in]{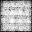}} &
\subfloat[epoch 50]{\includegraphics[width = 0.4in]{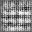}}&
\subfloat[epoch 65]{\includegraphics[width = 0.4in]{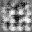}}&
\subfloat[epoch 80]{\includegraphics[width = 0.4in]{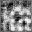}}&
\subfloat[epoch 95]{\includegraphics[width = 0.4in]{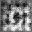}}&
\subfloat[epoch 110]{\includegraphics[width = 0.4in]{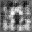}}&
\subfloat[epoch 125]{\includegraphics[width = 0.4in]{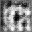}}&
\subfloat[epoch 140]{\includegraphics[width = 0.4in]{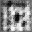}} &
\subfloat[epoch 155]{\includegraphics[width = 0.4in]{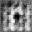}} \\
\subfloat[epoch 5]{\includegraphics[width = 0.4in]{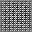}} &
\subfloat[epoch 20]{\includegraphics[width = 0.4in]{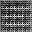}} &
\subfloat[epoch 35]{\includegraphics[width = 0.4in]{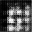}} &
\subfloat[epoch 50]{\includegraphics[width = 0.4in]{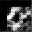}}&
\subfloat[epoch 65]{\includegraphics[width = 0.4in]{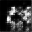}}&
\subfloat[epoch 80]{\includegraphics[width = 0.4in]{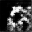}}&
\subfloat[epoch 95]{\includegraphics[width = 0.4in]{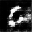}}&
\subfloat[epoch 110]{\includegraphics[width = 0.4in]{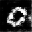}}&
\subfloat[epoch 125]{\includegraphics[width = 0.4in]{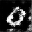}}&
\subfloat[epoch 140]{\includegraphics[width = 0.4in]{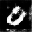}} &
\subfloat[epoch 155]{\includegraphics[width = 0.4in]{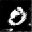}} 
\end{tabular}

\caption{DCGAN with No influence vs. influence in Collaborative Learning for 0 (Zero)}
\label{fig:DCGANbbb}
\end{figure*}

\begin{figure*}
\centering
\captionsetup[subfigure]{labelformat=empty, justification=centering}
\begin{tabular}{cccccccccccc}
\centering
\subfloat[epoch 5]{\includegraphics[width = 0.4in]{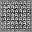}} &
\subfloat[epoch 20]{\includegraphics[width = 0.4in]{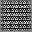}} &
\subfloat[epoch 35]{\includegraphics[width = 0.4in]{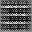}} &
\subfloat[epoch 50]{\includegraphics[width = 0.4in]{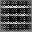}}&
\subfloat[epoch 65]{\includegraphics[width = 0.4in]{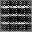}}&
\subfloat[epoch 80]{\includegraphics[width = 0.4in]{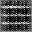}}&
\subfloat[epoch 95]{\includegraphics[width = 0.4in]{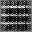}}&
\subfloat[epoch 110]{\includegraphics[width = 0.4in]{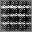}}&
\subfloat[epoch 125]{\includegraphics[width = 0.4in]{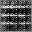}}&
\subfloat[epoch 140]{\includegraphics[width = 0.4in]{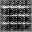}} &
\subfloat[epoch 155]{\includegraphics[width = 0.4in]{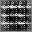}} \\
\subfloat[epoch 5]{\includegraphics[width = 0.4in]{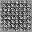}} &
\subfloat[epoch 20]{\includegraphics[width = 0.4in]{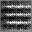}} &
\subfloat[epoch 35]{\includegraphics[width = 0.4in]{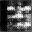}} &
\subfloat[epoch 50]{\includegraphics[width = 0.4in]{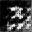}}&
\subfloat[epoch 65]{\includegraphics[width = 0.4in]{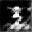}}&
\subfloat[epoch 80]{\includegraphics[width = 0.4in]{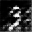}}&
\subfloat[epoch 95]{\includegraphics[width = 0.4in]{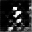}}&
\subfloat[epoch 110]{\includegraphics[width = 0.4in]{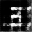}}&
\subfloat[epoch 125]{\includegraphics[width = 0.4in]{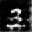}}&
\subfloat[epoch 140]{\includegraphics[width = 0.4in]{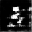}} &
\subfloat[epoch 155]{\includegraphics[width = 0.4in]{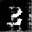}} 
\end{tabular}

\caption{DCGAN with No influence vs. influence in Collaborative Learning for 3 (Three)}
\label{fig:DCGANccc}
\end{figure*}

\subsubsection{Experiments on AT\&T} We performed similar experiments on the AT\&T dataset which consists of faces from 40 different people.  Initially, we tested the two-participant scenario, where one is the victim, and the other is the adversary. We assigned the first 20 classes to the first user and the remaining 20 classes to the adversary.  An extra class is given to the adversary to influence the training process. We ran several configurations with different upload rates, see Figure \ref{fig:OlivettiNoDP}.  The results show the adversary can get considerably good reconstructions of the targeted face. Some images are noisier than others, but this can hardly be improved given that the accuracy of the model tends to stay low for this particular dataset.

We have also implemented a multi-participant scenario, see Figure \ref{ppdl_architecture}, with 41 participants, 40 of which are honest and 1 is adversarial. Each honest participant possesses images pertaining to one class as training data, while the adversary has \emph{no} training data of his own. Namely, the adversary only trains on the images produced by the generator (G). The results (with $\theta_u = 1, \theta_d = 1$) are very good even when differential privacy is enabled (Figure \ref{ppdl_architecture}).

\begin{figure*}[!t]
\centering
\captionsetup[subfigure]{labelformat=empty, justification=centering}
\begin{tabular}{c|cccc}
\centering
\subfloat[Original]{\includegraphics[width = 0.6in]{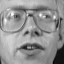}} &
\subfloat[$\frac{\epsilon}{c} = 100 \newline \theta_u = 1$]{\includegraphics[width = 0.6in]{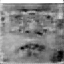}} &
\subfloat[$\frac{\epsilon}{c} = 100 \newline \theta_u = 0.1$]{\includegraphics[width = 0.6in]{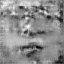}} &
\subfloat[$\frac{\epsilon}{c} = 10 \newline \theta_u = 1$]{\includegraphics[width = 0.6in]{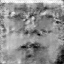}} &
\subfloat[$\frac{\epsilon}{c} = 10 \newline \theta_u = 0.1$]{\includegraphics[width = 0.6in]{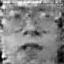}}
\end{tabular}

\caption{Experimental results on the AT\&T Dataset with 100\% download ($(\theta_d = 1)$ and DP enabled. Unlike MNIST, images are noisier because this particular dataset is small and the accuracy of the model is significantly affected when upload rates are small.}
\label{fig:OlivettiDP}
\end{figure*}

\begin{figure}
\centering
\begin{tabular}{cc}
\subfloat[]{\includegraphics[width = 0.5in]{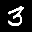}} &
\subfloat[]{\includegraphics[width = 0.5in]{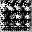}}
\end{tabular}
\caption{GAN Attack Results on the MNIST Dataset (left: original image, right: generated one) with DP Setup: $\frac{\epsilon}{c} = 0.01, \tau = 0.0001, \gamma = 0.001, \theta_u = 1, \theta_d = 1$. The value of $\epsilon$ is so small that the accuracy of the model does not increase. Since there is no learning, the GAN fails to produce clear results.}
\label{fig:TightPrivacyBounds}
\end{figure}

\begin{figure}
\centering
\begin{tabular}{cc}
\subfloat[]{\includegraphics[width = 0.5in]{original_11_man_glasses.png}} &
\subfloat[]{\includegraphics[width = 0.5in]{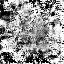}}
\end{tabular}
\caption{GAN Attack Results on the AT\&T Dataset (left: original image, right: generated one) with DP Setup: $\frac{\epsilon}{c} = 0.01, \tau = 0.0001, \gamma = 0.001, \theta_u = 1, \theta_d = 1$. The value of $\epsilon$ is so small that the accuracy of the model does not increase. Since there is no learning, the GAN fails to produce clear results.}
\label{fig:TightPrivacyBoundsATT}
\end{figure}

\begin{figure}
\centering
\setlength{\tabcolsep}{2pt}
\begin{tabular}{cc}
\setlength{\tabcolsep}{1pt}
\subfloat[$\frac{\epsilon}{c} = 100, \theta_u = 1, \theta_d = 1$]{
\begin{tabular}{lllll}

\includegraphics[width = 0.4in]{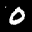}  & \includegraphics[width = 0.4in]{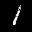}  & \includegraphics[width = 0.4in]{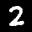}  & \includegraphics[width = 0.4in]{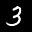}  & \includegraphics[width = 0.4in]{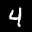} \\ 
\includegraphics[width = 0.4in]{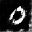}  & \includegraphics[width = 0.4in]{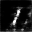}  & \includegraphics[width = 0.4in]{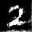}  & \includegraphics[width = 0.4in]{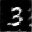}  & \includegraphics[width = 0.4in]{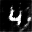}  \\
\end{tabular}
}
   \\
 \setlength{\tabcolsep}{1pt}
\subfloat[$\frac{\epsilon}{c} = 100, \theta_u = 0.1, \theta_d = 1$]{
\begin{tabular}{lllll}

\includegraphics[width = 0.4in]{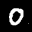}  & \includegraphics[width = 0.4in]{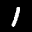}  & \includegraphics[width = 0.4in]{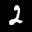}  & \includegraphics[width = 0.4in]{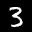}  & \includegraphics[width = 0.4in]{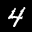} \\ 
\includegraphics[width = 0.4in]{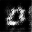}  & \includegraphics[width = 0.4in]{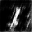}  & \includegraphics[width = 0.4in]{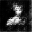}  & \includegraphics[width = 0.4in]{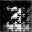}  & \includegraphics[width = 0.4in]{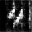}  \\
\end{tabular}
}

\\
\setlength{\tabcolsep}{1pt}
\subfloat[$\frac{\epsilon}{c} = 10, \theta_u = 1, \theta_d = 1$]{
\begin{tabular}{lllll}

\includegraphics[width = 0.4in]{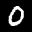}  & \includegraphics[width = 0.4in]{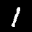}  & \includegraphics[width = 0.4in]{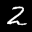}  & \includegraphics[width = 0.4in]{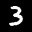}  & \includegraphics[width = 0.4in]{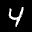} \\ 
\includegraphics[width = 0.4in]{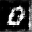}  & \includegraphics[width = 0.4in]{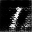}  & \includegraphics[width = 0.4in]{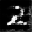}  & \includegraphics[width = 0.4in]{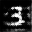}  & \includegraphics[width = 0.4in]{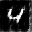}  \\
\end{tabular}
}
    \\
 \setlength{\tabcolsep}{1pt}
\subfloat[$\frac{\epsilon}{c} = 10, \theta_u = 0.1, \theta_d = 1$]{
\begin{tabular}{lllll}

\includegraphics[width = 0.4in]{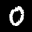}  & \includegraphics[width = 0.4in]{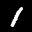}  & \includegraphics[width = 0.4in]{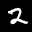}  & \includegraphics[width = 0.4in]{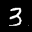}  & \includegraphics[width = 0.4in]{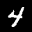} \\ 
\includegraphics[width = 0.4in]{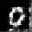}  & \includegraphics[width = 0.4in]{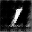}  & \includegraphics[width = 0.4in]{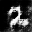}  & \includegraphics[width = 0.4in]{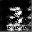}  & \includegraphics[width = 0.4in]{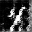}  \\
\end{tabular}
}
\end{tabular}
\caption{Results for the GAN attack on a two-user scenario with Differential Privacy enabled. Bottom row, samples generated by the GAN. Top row, samples from the training set closest to the ones generated by the GAN.}
\label{gandp}
\end{figure}

\subsection{GAN Attack, No Influence vs. Influence on Collaborative Learning}

One may wonder about the effect of the fake label to the collaborative learning. Recall that images generated by the generative model are placed into an artificial class to trick the victim into releasing finer details on the targeted class. We measured the effect of the adversarial influence, and we experimentally confirmed that its effect is remarkable: The learning gets faster, but also the information retrieved by the adversary is significantly better.
We ran the experiments until the accuracy of the model on the testing set was above $97\%$, collaboratively training a CNN model. The datasets of both the adversary and the victim are separated from each other, and there are no labels in common.

In Figures \ref{fig:DCGANbbb} and \ref{fig:DCGANccc}, we show the result of the passive GAN attack with the standard GAN attack proposed in Section \ref{attack}, when we are trying to recover, respectively, 0's and 3's from the first user. In the top row, we show the images from the passive attack with no influence and in the bottom row the images from the standard procedure with the influence of the artificial class. 
The effect of the adversarial influence is evident, and images appear much clearer and crisper even after only 50 epochs per participant. During our experiments, we noticed that $G$ starts producing good results as soon as the accuracy of the model reaches 80\%.

\subsection{GAN Attack on Differentially Private Collaborative Learning}

It has been argued in \cite{shokriPPDL} that differential privacy can be used to add noise to the parameters of the deep learning model {\em ``to ensure that parameter updates do not leak too much information about any individual point in the training dataset."} (Quoted from \cite{shokriPPDL}.)
The authors consider only a {\em passive} adversary and rely on differential privacy to mitigate possible leakages that might come from parameter updates. They highlight two cases of potential leakage: (i) the way how gradient selection is performed and (ii) actual values of the shared gradients. To address both of these issues, the approach in \cite{shokriPPDL} relies on sparse vector technique \cite{dwork2014algorithmic}. For each epoch (iteration) of the collaborative learning process, they define a total privacy budget $\epsilon$ for each participant. This budget is split into $c$ parts, where $c$ is the total number of gradients that can be shared per epoch. A portion of gradients is randomly select such that they are above a threshold ($\tau$). They dedicate $\frac {8}{9}$ of $\frac {\epsilon} {c}$ to the selection of the parameters and use the remaining $\frac {1}{9}$ to release the value. They rely on the Laplacian mechanism to add noise during selection as well as sharing of the parameters, in agreement with the allocated privacy budget.

To demonstrate that record-level differential privacy is ineffective against an active adversary, we ran the collaborative learning process between the two participants ($A$ and $V$) with differential privacy enabled. We kept the datasets of the participants distinct: In MNIST experiments, $V$ had only records of classes from 0 to 4 and $A$ had records of classes from 5 to 9 plus the artificial class that $A$ introduces. For the AT\&T experiments, $V$ has records for the first 20 classes in the dataset and $A$ for the next 20 classes plus the artificial class as in Subsection \ref{ganattack}. During our experiments we kept the download rate ($\theta_d$) fixed at 100\%, threshold ($\tau$) at 0.0001 and the range ($\gamma$) at 0.001, similar to \cite{shokriPPDL}. On Figures  \ref{fig:OlivettiDP} and \ref{gandp}, we provide results for a privacy budget per parameter ($\frac{\epsilon}{c}$) of 100 and 10 and varying upload rate ($\theta_u$). Even though it takes longer for the models to converge under the differential privacy constraints, our results demonstrate our claim, i.e., \emph{as long as the training process is successful and the model is converging, $G$ can generate good results}. 

\paragraph{On the $\epsilon$ value.} We observe that the $\epsilon$ in \cite{shokriPPDL} is very large and the effect of differential privacy may be questionable. However, with small $\epsilon$, the local models are unable to learn and collaborative learning fails completely. This is consistent with what is reported in  \cite{shokriPPDL}. Indeed, we ran our experiments with tighter privacy constraints. The generator failed to produce good results but because the local model were unable to learn at all. In Figure \ref{fig:TightPrivacyBounds} and \ref{fig:TightPrivacyBoundsATT} we show an example where we set a tighter privacy bound, which translates into stronger differential privacy guarantees, and the GAN is ineffective. At the same time, this is expected since the local model and the one in the parameter server are unable to learn and collaborative learning is not happening. 
It is possible to use the techniques in \cite{goodfellowDeepLearningwithDP} to bring $\epsilon$ down to a single-digit value. However, we stress again that our attack is independent of whatever record-level DP implementation is used. The GAN will generate good samples as long as the discriminator is learning (see Figure \ref{victim_adversary_illustration}).

\section{Conclusions}
\label{conclusions}
In this work, we propose and implement a novel class of active inference attacks on deep neural networks in a collaborative setting. Our approach relies on Generative Adversarial Networks (GANs) and is more effective and general than existing information extraction mechanisms. We believe our work will have a significant impact in the real world as major companies are considering distributed, federated, or decentralized deep learning approaches to protect the privacy of users. 

The main point of our research is that collaborative learning is less desirable than the centralized learning approach it is supposed to replace. In collaborative learning, any user may violate the privacy of other users in the system without involving the service provider. 

Finally, we were not able to devise effective countermeasures against our attack. Solutions may rely on secure multiparty computation or (fully) homomorphic encryption. However:  (1) privacy-preserving collaborative learning was introduced as a way to avoid these costly cryptographic primitives \cite{shokriPPDL}, and (2) the solutions we explored based on them would still be susceptible to some forms of our attack. 
Another approach is to consider differential privacy at different granularities. User or device-level DP would protect against the attacks devised in this paper. However, it's not clear yet how to build a real system for collaborative learning with device, class, or user-level DP (e.g., users behave and share data in unpredictable ways). Therefore, we leave this subject for future work. 

\section*{Acknowledgment}
We thank Mart\'{i}n Abadi, Matt Fredrikson, Thomas Ristenpart, Vitaly Shmatikov, and Adam Smith for their insightful comments that greatly improved our paper. 
We are grateful to the authors of \cite{shokriPPDL} for providing us with the source code of their implementation of privacy-preserving collaborative deep learning.

\bibliographystyle{ACM-Reference-Format}
\balance
\bibliography{ccs17_main}


\begin{thebibliography}{00}


\ifx \showCODEN    \undefined \def \showCODEN     #1{\unskip}     \fi
\ifx \showDOI      \undefined \def \showDOI       #1{#1}\fi
\ifx \showISBNx    \undefined \def \showISBNx     #1{\unskip}     \fi
\ifx \showISBNxiii \undefined \def \showISBNxiii  #1{\unskip}     \fi
\ifx \showISSN     \undefined \def \showISSN      #1{\unskip}     \fi
\ifx \showLCCN     \undefined \def \showLCCN      #1{\unskip}     \fi
\ifx \shownote     \undefined \def \shownote      #1{#1}          \fi
\ifx \showarticletitle \undefined \def \showarticletitle #1{#1}   \fi
\ifx \showURL      \undefined \def \showURL       {\relax}        \fi
\providecommand\bibfield[2]{#2}
\providecommand\bibinfo[2]{#2}
\providecommand\natexlab[1]{#1}
\providecommand\showeprint[2][]{arXiv:#2}

\bibitem[\protect\citeauthoryear{Abadi, Chu, Goodfellow, McMahan, Mironov,
  Talwar, and Zhang}{Abadi et~al\mbox{.}}{2016}]%
        {goodfellowDeepLearningwithDP}
\bibfield{author}{\bibinfo{person}{Mart{\'\i}n Abadi}, \bibinfo{person}{Andy
  Chu}, \bibinfo{person}{Ian Goodfellow}, \bibinfo{person}{H~Brendan McMahan},
  \bibinfo{person}{Ilya Mironov}, \bibinfo{person}{Kunal Talwar}, {and}
  \bibinfo{person}{Li Zhang}.} \bibinfo{year}{2016}\natexlab{}.
\newblock \showarticletitle{Deep learning with differential privacy}. In
  \bibinfo{booktitle}{{\em Proceedings of the 2016 ACM SIGSAC Conference on
  Computer and Communications Security}}. ACM, \bibinfo{pages}{308--318}.
\newblock


\bibitem[\protect\citeauthoryear{Abdulkader, Lakshmiratan, and
  Zhang}{Abdulkader et~al\mbox{.}}{2016}]%
        {DeepTextFCB}
\bibfield{author}{\bibinfo{person}{Ahmad Abdulkader}, \bibinfo{person}{Aparna
  Lakshmiratan}, {and} \bibinfo{person}{Joy Zhang}.}
  \bibinfo{year}{2016}\natexlab{}.
\newblock \bibinfo{title}{Introducing DeepText: Facebook's text understanding
  engine}.
\newblock   (\bibinfo{year}{2016}).
\newblock
\showURL{%
\url{https://tinyurl.com/jj359dv}}


\bibitem[\protect\citeauthoryear{Arjovsky and Bottou}{Arjovsky and
  Bottou}{2017}]%
        {Bottou17}
\bibfield{author}{\bibinfo{person}{Martin Arjovsky} {and}
  \bibinfo{person}{L{\'e}on Bottou}.} \bibinfo{year}{2017}\natexlab{}.
\newblock \showarticletitle{Towards principled methods for training generative
  adversarial networks}. In \bibinfo{booktitle}{{\em 5th International
  Conference on Learning Representations (ICLR)}}.
\newblock


\bibitem[\protect\citeauthoryear{Ateniese, Mancini, Spognardi, Villani, Vitali,
  and Felici}{Ateniese et~al\mbox{.}}{2015}]%
        {ateniese2015hacking}
\bibfield{author}{\bibinfo{person}{Giuseppe Ateniese}, \bibinfo{person}{Luigi~V
  Mancini}, \bibinfo{person}{Angelo Spognardi}, \bibinfo{person}{Antonio
  Villani}, \bibinfo{person}{Domenico Vitali}, {and} \bibinfo{person}{Giovanni
  Felici}.} \bibinfo{year}{2015}\natexlab{}.
\newblock \showarticletitle{Hacking smart machines with smarter ones: How to
  extract meaningful data from machine learning classifiers}.
\newblock \bibinfo{journal}{{\em International Journal of Security and
  Networks\/}} \bibinfo{volume}{10}, \bibinfo{number}{3}
  (\bibinfo{year}{2015}), \bibinfo{pages}{137--150}.
\newblock
\showURL{%
\url{https://arxiv.org/abs/1306.4447}}


\bibitem[\protect\citeauthoryear{Barthe, Fong, Gaboardi, Gr{\'e}goire, Hsu, and
  Strub}{Barthe et~al\mbox{.}}{2016}]%
        {bartheAdvancedDP}
\bibfield{author}{\bibinfo{person}{Gilles Barthe}, \bibinfo{person}{No{\'e}mie
  Fong}, \bibinfo{person}{Marco Gaboardi}, \bibinfo{person}{Benjamin
  Gr{\'e}goire}, \bibinfo{person}{Justin Hsu}, {and}
  \bibinfo{person}{Pierre-Yves Strub}.} \bibinfo{year}{2016}\natexlab{}.
\newblock \showarticletitle{Advanced probabilistic couplings for differential
  privacy}. In \bibinfo{booktitle}{{\em Proceedings of the 2016 ACM SIGSAC
  Conference on Computer and Communications Security}}. ACM,
  \bibinfo{pages}{55--67}.
\newblock


\bibitem[\protect\citeauthoryear{Bengio}{Bengio}{2009}]%
        {BengioLearningDeepArchitectures}
\bibfield{author}{\bibinfo{person}{Yoshua Bengio}.}
  \bibinfo{year}{2009}\natexlab{}.
\newblock \showarticletitle{Learning Deep Architectures for AI}.
\newblock \bibinfo{journal}{{\em Found. Trends Mach. Learn.\/}}
  \bibinfo{volume}{2}, \bibinfo{number}{1} (\bibinfo{date}{Jan.}
  \bibinfo{year}{2009}), \bibinfo{pages}{1--127}.
\newblock
\showISSN{1935-8237}
\showDOI{%
\url{https://doi.org/10.1561/2200000006}}


\bibitem[\protect\citeauthoryear{Blocki, Datta, and Bonneau}{Blocki
  et~al\mbox{.}}{2016}]%
        {blocki2016differentially}
\bibfield{author}{\bibinfo{person}{Jeremiah Blocki}, \bibinfo{person}{Anupam
  Datta}, {and} \bibinfo{person}{Joseph Bonneau}.}
  \bibinfo{year}{2016}\natexlab{}.
\newblock \showarticletitle{Differentially Private Password Frequency Lists}.
  In \bibinfo{booktitle}{{\em NDSS'16}}.
\newblock


\bibitem[\protect\citeauthoryear{Bonawitz, Ivanov, Kreuter, Marcedone, McMahan,
  Patel, Ramage, Segal, and Seth}{Bonawitz et~al\mbox{.}}{2017}]%
        {bonawitzpractical}
\bibfield{author}{\bibinfo{person}{Keith Bonawitz}, \bibinfo{person}{Vladimir
  Ivanov}, \bibinfo{person}{Ben Kreuter}, \bibinfo{person}{Antonio Marcedone},
  \bibinfo{person}{H~Brendan McMahan}, \bibinfo{person}{Sarvar Patel},
  \bibinfo{person}{Daniel Ramage}, \bibinfo{person}{Aaron Segal}, {and}
  \bibinfo{person}{Karn Seth}.} \bibinfo{year}{2017}\natexlab{}.
\newblock \showarticletitle{Practical Secure Aggregation for Privacy Preserving
  Machine Learning}.
\newblock  (\bibinfo{year}{2017}).
\newblock


\bibitem[\protect\citeauthoryear{Bouchacourt, Mudigonda, and
  Nowozin}{Bouchacourt et~al\mbox{.}}{2016}]%
        {Bouchacourt16}
\bibfield{author}{\bibinfo{person}{Diane Bouchacourt}, \bibinfo{person}{Pawan~K
  Mudigonda}, {and} \bibinfo{person}{Sebastian Nowozin}.}
  \bibinfo{year}{2016}\natexlab{}.
\newblock \showarticletitle{DISCO Nets: DISsimilarity COefficients Networks}.
  In \bibinfo{booktitle}{{\em Advances in Neural Information Processing
  Systems}}. \bibinfo{pages}{352--360}.
\newblock


\bibitem[\protect\citeauthoryear{Breiman}{Breiman}{2001}]%
        {Breiman01}
\bibfield{author}{\bibinfo{person}{Leo Breiman}.}
  \bibinfo{year}{2001}\natexlab{}.
\newblock \showarticletitle{Random Forests}.
\newblock \bibinfo{journal}{{\em Machine Learning\/}} \bibinfo{volume}{45},
  \bibinfo{number}{1} (\bibinfo{year}{2001}), \bibinfo{pages}{5--32}.
\newblock


\bibitem[\protect\citeauthoryear{Bun and Steinke}{Bun and Steinke}{2016}]%
        {bun2016concentrated}
\bibfield{author}{\bibinfo{person}{Mark Bun} {and} \bibinfo{person}{Thomas
  Steinke}.} \bibinfo{year}{2016}\natexlab{}.
\newblock \showarticletitle{Concentrated differential privacy: Simplifications,
  extensions, and lower bounds}. In \bibinfo{booktitle}{{\em Theory of
  Cryptography Conference}}. Springer, \bibinfo{pages}{635--658}.
\newblock


\bibitem[\protect\citeauthoryear{Camenisch, Manulis, Tsudik, and
  Wright}{Camenisch et~al\mbox{.}}{2012}]%
        {Camenisch2012PrivacyOrientedC}
\bibfield{author}{\bibinfo{person}{Jan Camenisch}, \bibinfo{person}{Mark
  Manulis}, \bibinfo{person}{Gene Tsudik}, {and} \bibinfo{person}{Rebecca~N.
  Wright}.} \bibinfo{year}{2012}\natexlab{}.
\newblock \showarticletitle{Privacy-Oriented Cryptography (Dagstuhl Seminar
  12381)}.
\newblock \bibinfo{journal}{{\em Dagstuhl Reports\/}}  \bibinfo{volume}{2}
  (\bibinfo{year}{2012}), \bibinfo{pages}{165--183}.
\newblock
\showURL{%
\url{http://drops.dagstuhl.de/opus/volltexte/2013/3755/pdf/dagrep_v002_i009_p165_s12381.pdf}}


\bibitem[\protect\citeauthoryear{Chaudhuri, Monteleoni, and Sarwate}{Chaudhuri
  et~al\mbox{.}}{2011}]%
        {chaudhuri2011differentially}
\bibfield{author}{\bibinfo{person}{Kamalika Chaudhuri}, \bibinfo{person}{Claire
  Monteleoni}, {and} \bibinfo{person}{Anand~D Sarwate}.}
  \bibinfo{year}{2011}\natexlab{}.
\newblock \showarticletitle{Differentially private empirical risk
  minimization}.
\newblock \bibinfo{journal}{{\em Journal of machine learning research: JMLR\/}}
   \bibinfo{volume}{12} (\bibinfo{year}{2011}), \bibinfo{pages}{1069}.
\newblock


\bibitem[\protect\citeauthoryear{Chintala}{Chintala}{2016}]%
        {chintalaTorchCode}
\bibfield{author}{\bibinfo{person}{Soumith Chintala}.}
  \bibinfo{year}{2016}\natexlab{}.
\newblock \bibinfo{title}{DCGAN.torch: Train your own image generator}.
\newblock   (\bibinfo{year}{2016}).
\newblock
\showURL{%
\url{https://github.com/soumith/dcgan.torch}}


\bibitem[\protect\citeauthoryear{Collobert, Weston, Bottou, Karlen,
  Kavukcuoglu, and Kuksa}{Collobert et~al\mbox{.}}{2011}]%
        {collobert2011natural}
\bibfield{author}{\bibinfo{person}{Ronan Collobert}, \bibinfo{person}{Jason
  Weston}, \bibinfo{person}{L{\'e}on Bottou}, \bibinfo{person}{Michael Karlen},
  \bibinfo{person}{Koray Kavukcuoglu}, {and} \bibinfo{person}{Pavel Kuksa}.}
  \bibinfo{year}{2011}\natexlab{}.
\newblock \showarticletitle{Natural language processing (almost) from scratch}.
\newblock \bibinfo{journal}{{\em Journal of Machine Learning Research\/}}
  \bibinfo{volume}{12}, \bibinfo{number}{Aug} (\bibinfo{year}{2011}),
  \bibinfo{pages}{2493--2537}.
\newblock


\bibitem[\protect\citeauthoryear{Cruz-Roa, Ovalle, Madabhushi, and
  Osorio}{Cruz-Roa et~al\mbox{.}}{2013}]%
        {cruz2013deep}
\bibfield{author}{\bibinfo{person}{Angel~Alfonso Cruz-Roa},
  \bibinfo{person}{John Edison~Arevalo Ovalle}, \bibinfo{person}{Anant
  Madabhushi}, {and} \bibinfo{person}{Fabio Augusto~Gonz{\'a}lez Osorio}.}
  \bibinfo{year}{2013}\natexlab{}.
\newblock \showarticletitle{A deep learning architecture for image
  representation, visual interpretability and automated basal-cell carcinoma
  cancer detection}. In \bibinfo{booktitle}{{\em International Conference on
  Medical Image Computing and Computer-Assisted Intervention}}. Springer Berlin
  Heidelberg, \bibinfo{pages}{403--410}.
\newblock


\bibitem[\protect\citeauthoryear{Dean, Corrado, Monga, Chen, Devin, Mao,
  Senior, Tucker, Yang, Le, et~al\mbox{.}}{Dean et~al\mbox{.}}{2012}]%
        {dean2012large}
\bibfield{author}{\bibinfo{person}{Jeffrey Dean}, \bibinfo{person}{Greg
  Corrado}, \bibinfo{person}{Rajat Monga}, \bibinfo{person}{Kai Chen},
  \bibinfo{person}{Matthieu Devin}, \bibinfo{person}{Mark Mao},
  \bibinfo{person}{Andrew Senior}, \bibinfo{person}{Paul Tucker},
  \bibinfo{person}{Ke Yang}, \bibinfo{person}{Quoc~V Le}, {et~al\mbox{.}}}
  \bibinfo{year}{2012}\natexlab{}.
\newblock \showarticletitle{Large scale distributed deep networks}. In
  \bibinfo{booktitle}{{\em Advances in neural information processing systems}}.
  \bibinfo{pages}{1223--1231}.
\newblock


\bibitem[\protect\citeauthoryear{DeepMind}{DeepMind}{2016}]%
        {deepmindHealthcare}
\bibfield{author}{\bibinfo{person}{DeepMind}.} \bibinfo{year}{2016}\natexlab{}.
\newblock \bibinfo{title}{DeepMind Health, Clinician-led. Patient-centred.}
\newblock   (\bibinfo{year}{2016}).
\newblock
\showURL{%
\url{https://deepmind.com/applied/deepmind-health/}}


\bibitem[\protect\citeauthoryear{Diakonikolas, Hardt, and Schmidt}{Diakonikolas
  et~al\mbox{.}}{2015}]%
        {diakonikolas}
\bibfield{author}{\bibinfo{person}{Ilias Diakonikolas}, \bibinfo{person}{Moritz
  Hardt}, {and} \bibinfo{person}{Ludwig Schmidt}.}
  \bibinfo{year}{2015}\natexlab{}.
\newblock \showarticletitle{Differentially private learning of structured
  discrete distributions}. In \bibinfo{booktitle}{{\em Advances in Neural
  Information Processing Systems}}. \bibinfo{pages}{2566--2574}.
\newblock


\bibitem[\protect\citeauthoryear{Dowlin, Gilad-Bachrach, Laine, Lauter,
  Michael, and Wernsing}{Dowlin et~al\mbox{.}}{2016}]%
        {CryptoNets:Microsoft}
\bibfield{author}{\bibinfo{person}{Nathan Dowlin}, \bibinfo{person}{Ran
  Gilad-Bachrach}, \bibinfo{person}{Kim Laine}, \bibinfo{person}{Kristin
  Lauter}, \bibinfo{person}{Naehrig Michael}, {and} \bibinfo{person}{John
  Wernsing}.} \bibinfo{year}{2016}\natexlab{}.
\newblock \bibinfo{booktitle}{{\em CryptoNets: Applying Neural Networks to
  Encrypted Data with High Throughput and Accuracy}}.
\newblock \bibinfo{type}{{T}echnical {R}eport} MSR-TR-2016-3.
\newblock
\showURL{%
\url{http://research.microsoft.com/apps/pubs/default.aspx?id=260989}}


\bibitem[\protect\citeauthoryear{Dwork}{Dwork}{2006}]%
        {dwork2006differential}
\bibfield{author}{\bibinfo{person}{Cynthia Dwork}.}
  \bibinfo{year}{2006}\natexlab{}.
\newblock \showarticletitle{Differential privacy}.
\newblock In \bibinfo{booktitle}{{\em Automata, Languages and Programming, 33rd
  International Colloquium, {ICALP} 2006, Venice, Italy, July 10-14, 2006,
  Proceedings, Part {II}}}. \bibinfo{publisher}{Springer Berlin Heidelberg},
  \bibinfo{pages}{1--12}.
\newblock


\bibitem[\protect\citeauthoryear{Dwork and Naor}{Dwork and Naor}{2008}]%
        {dwork2008difficulties}
\bibfield{author}{\bibinfo{person}{Cynthia Dwork} {and} \bibinfo{person}{Moni
  Naor}.} \bibinfo{year}{2008}\natexlab{}.
\newblock \showarticletitle{On the difficulties of disclosure prevention in
  statistical databases or the case for differential privacy}.
\newblock \bibinfo{journal}{{\em Journal of Privacy and Confidentiality\/}}
  \bibinfo{volume}{2}, \bibinfo{number}{1} (\bibinfo{year}{2008}),
  \bibinfo{pages}{8}.
\newblock


\bibitem[\protect\citeauthoryear{Dwork and Roth}{Dwork and Roth}{2014}]%
        {dwork2014algorithmic}
\bibfield{author}{\bibinfo{person}{Cynthia Dwork} {and} \bibinfo{person}{Aaron
  Roth}.} \bibinfo{year}{2014}\natexlab{}.
\newblock \showarticletitle{The algorithmic foundations of differential
  privacy}.
\newblock \bibinfo{journal}{{\em Foundations and Trends in Theoretical Computer
  Science\/}} \bibinfo{volume}{9}, \bibinfo{number}{3-4}
  (\bibinfo{year}{2014}), \bibinfo{pages}{211--407}.
\newblock


\bibitem[\protect\citeauthoryear{Dwork and Rothblum}{Dwork and
  Rothblum}{2016}]%
        {dwork2016concentrated}
\bibfield{author}{\bibinfo{person}{Cynthia Dwork} {and} \bibinfo{person}{Guy~N
  Rothblum}.} \bibinfo{year}{2016}\natexlab{}.
\newblock \showarticletitle{Concentrated differential privacy}.
\newblock \bibinfo{journal}{{\em arXiv preprint arXiv:1603.01887\/}}
  (\bibinfo{year}{2016}).
\newblock


\bibitem[\protect\citeauthoryear{Eigner, Kate, Maffei, Pampaloni, and
  Pryvalov}{Eigner et~al\mbox{.}}{2014}]%
        {eigner2014differentially}
\bibfield{author}{\bibinfo{person}{Fabienne Eigner}, \bibinfo{person}{Aniket
  Kate}, \bibinfo{person}{Matteo Maffei}, \bibinfo{person}{Francesca
  Pampaloni}, {and} \bibinfo{person}{Ivan Pryvalov}.}
  \bibinfo{year}{2014}\natexlab{}.
\newblock \showarticletitle{Differentially private data aggregation with
  optimal utility}. In \bibinfo{booktitle}{{\em Proceedings of the 30th Annual
  Computer Security Applications Conference}}. ACM, \bibinfo{pages}{316--325}.
\newblock


\bibitem[\protect\citeauthoryear{Fakoor, Ladhak, Nazi, and Huber}{Fakoor
  et~al\mbox{.}}{2013}]%
        {fakoor2013using}
\bibfield{author}{\bibinfo{person}{Rasool Fakoor}, \bibinfo{person}{Faisal
  Ladhak}, \bibinfo{person}{Azade Nazi}, {and} \bibinfo{person}{Manfred
  Huber}.} \bibinfo{year}{2013}\natexlab{}.
\newblock \showarticletitle{Using deep learning to enhance cancer diagnosis and
  classification}. In \bibinfo{booktitle}{{\em The 30th International
  Conference on Machine Learning (ICML 2013),WHEALTH workshop}}.
\newblock


\bibitem[\protect\citeauthoryear{Fredrikson, Jha, and Ristenpart}{Fredrikson
  et~al\mbox{.}}{2015}]%
        {fredriksonMIAttack}
\bibfield{author}{\bibinfo{person}{Matt Fredrikson}, \bibinfo{person}{Somesh
  Jha}, {and} \bibinfo{person}{Thomas Ristenpart}.}
  \bibinfo{year}{2015}\natexlab{}.
\newblock \showarticletitle{Model inversion attacks that exploit confidence
  information and basic countermeasures}. In \bibinfo{booktitle}{{\em
  Proceedings of the 22nd ACM SIGSAC Conference on Computer and Communications
  Security}}. ACM, \bibinfo{pages}{1322--1333}.
\newblock


\bibitem[\protect\citeauthoryear{Fredrikson, Lantz, Jha, Lin, Page, and
  Ristenpart}{Fredrikson et~al\mbox{.}}{2014}]%
        {fredrikson2014privacy}
\bibfield{author}{\bibinfo{person}{Matthew Fredrikson}, \bibinfo{person}{Eric
  Lantz}, \bibinfo{person}{Somesh Jha}, \bibinfo{person}{Simon Lin},
  \bibinfo{person}{David Page}, {and} \bibinfo{person}{Thomas Ristenpart}.}
  \bibinfo{year}{2014}\natexlab{}.
\newblock \showarticletitle{Privacy in pharmacogenetics: An end-to-end case
  study of personalized warfarin dosing}. In \bibinfo{booktitle}{{\em 23rd
  USENIX Security Symposium (USENIX Security 14)}}. \bibinfo{pages}{17--32}.
\newblock


\bibitem[\protect\citeauthoryear{Goodfellow, Bengio, and Courville}{Goodfellow
  et~al\mbox{.}}{2016}]%
        {Goodfellow-et-al-2016-Book}
\bibfield{author}{\bibinfo{person}{Ian Goodfellow}, \bibinfo{person}{Yoshua
  Bengio}, {and} \bibinfo{person}{Aaron Courville}.}
  \bibinfo{year}{2016}\natexlab{}.
\newblock \bibinfo{booktitle}{{\em Deep learning}}.
\newblock \bibinfo{publisher}{MIT Press}.
\newblock


\bibitem[\protect\citeauthoryear{Goodfellow, Pouget-Abadie, Mirza, Xu,
  Warde-Farley, Ozair, Courville, and Bengio}{Goodfellow et~al\mbox{.}}{2014}]%
        {NIPS2014_5423}
\bibfield{author}{\bibinfo{person}{Ian Goodfellow}, \bibinfo{person}{Jean
  Pouget-Abadie}, \bibinfo{person}{Mehdi Mirza}, \bibinfo{person}{Bing Xu},
  \bibinfo{person}{David Warde-Farley}, \bibinfo{person}{Sherjil Ozair},
  \bibinfo{person}{Aaron Courville}, {and} \bibinfo{person}{Yoshua Bengio}.}
  \bibinfo{year}{2014}\natexlab{}.
\newblock \showarticletitle{Generative adversarial nets}.
\newblock In \bibinfo{booktitle}{{\em Advances in neural information processing
  systems}}. \bibinfo{pages}{2672--2680}.
\newblock


\bibitem[\protect\citeauthoryear{Goodfellow, Shlens, and Szegedy}{Goodfellow
  et~al\mbox{.}}{2015}]%
        {Shlens15}
\bibfield{author}{\bibinfo{person}{Ian Goodfellow}, \bibinfo{person}{Jonathon
  Shlens}, {and} \bibinfo{person}{Christian Szegedy}.}
  \bibinfo{year}{2015}\natexlab{}.
\newblock \showarticletitle{Explaining and Harnessing Adversarial Examples}. In
  \bibinfo{booktitle}{{\em International Conference on Learning
  Representations}}.
\newblock
\showURL{%
\url{https://arxiv.org/pdf/1412.6572v3.pdf}}


\bibitem[\protect\citeauthoryear{Goodfellow}{Goodfellow}{2014}]%
        {Goodfellow15}
\bibfield{author}{\bibinfo{person}{Ian~J Goodfellow}.}
  \bibinfo{year}{2014}\natexlab{}.
\newblock \showarticletitle{On distinguishability criteria for estimating
  generative models}.
\newblock \bibinfo{journal}{{\em arXiv preprint arXiv:1412.6515\/}}
  (\bibinfo{year}{2014}).
\newblock


\bibitem[\protect\citeauthoryear{{Google DeepMind}}{{Google DeepMind}}{2016}]%
        {googleGoSedol}
\bibfield{author}{\bibinfo{person}{{Google DeepMind}}.}
  \bibinfo{year}{2016}\natexlab{}.
\newblock \bibinfo{title}{AlphaGo, the first computer program to ever beat a
  professional player at the game of {GO}}.
\newblock   (\bibinfo{year}{2016}).
\newblock
\showURL{%
\url{https://deepmind.com/alpha-go}}


\bibitem[\protect\citeauthoryear{Graves, Mohamed, and Hinton}{Graves
  et~al\mbox{.}}{2013}]%
        {graves2013speech}
\bibfield{author}{\bibinfo{person}{Alex Graves}, \bibinfo{person}{Abdel-rahman
  Mohamed}, {and} \bibinfo{person}{Geoffrey Hinton}.}
  \bibinfo{year}{2013}\natexlab{}.
\newblock \showarticletitle{Speech recognition with deep recurrent neural
  networks}. In \bibinfo{booktitle}{{\em 2013 IEEE international conference on
  acoustics, speech and signal processing}}. IEEE, \bibinfo{pages}{6645--6649}.
\newblock


\bibitem[\protect\citeauthoryear{Greenberg}{Greenberg}{2016}]%
        {AppleDifferentialPrivacy}
\bibfield{author}{\bibinfo{person}{Andy Greenberg}.}
  \bibinfo{year}{2016}\natexlab{}.
\newblock \bibinfo{title}{Apple's 'Differential Privacy' Is About Collecting
  Your Data---But Not Your Data}.
\newblock   (\bibinfo{year}{2016}).
\newblock
\showURL{%
\url{https://www.wired.com/2016/06/apples-differential-privacy-collecting-data/}}


\bibitem[\protect\citeauthoryear{Grosse, Papernot, Manoharan, Backes, and
  McDaniel}{Grosse et~al\mbox{.}}{2016}]%
        {grosse2016adversarial}
\bibfield{author}{\bibinfo{person}{Kathrin Grosse}, \bibinfo{person}{Nicolas
  Papernot}, \bibinfo{person}{Praveen Manoharan}, \bibinfo{person}{Michael
  Backes}, {and} \bibinfo{person}{Patrick McDaniel}.}
  \bibinfo{year}{2016}\natexlab{}.
\newblock \showarticletitle{Adversarial Perturbations Against Deep Neural
  Networks for Malware Classification}.
\newblock \bibinfo{journal}{{\em arXiv preprint arXiv:1606.04435\/}}
  (\bibinfo{year}{2016}).
\newblock


\bibitem[\protect\citeauthoryear{Haeberlen, Pierce, and Narayan}{Haeberlen
  et~al\mbox{.}}{2011}]%
        {dpOnFire}
\bibfield{author}{\bibinfo{person}{Andreas Haeberlen},
  \bibinfo{person}{Benjamin~C. Pierce}, {and} \bibinfo{person}{Arjun Narayan}.}
  \bibinfo{year}{2011}\natexlab{}.
\newblock \showarticletitle{Differential Privacy Under Fire}. In
  \bibinfo{booktitle}{{\em Proceedings of the 20th USENIX Conference on
  Security}} {\em (\bibinfo{series}{SEC'11})}. \bibinfo{publisher}{USENIX
  Association}, \bibinfo{address}{Berkeley, CA, USA}, \bibinfo{pages}{33--33}.
\newblock
\showURL{%
\url{http://dl.acm.org/citation.cfm?id=2028067.2028100}}


\bibitem[\protect\citeauthoryear{Jain, Kulkarni, Thakurta, and Williams}{Jain
  et~al\mbox{.}}{2015}]%
        {jain2015drop}
\bibfield{author}{\bibinfo{person}{Prateek Jain}, \bibinfo{person}{Vivek
  Kulkarni}, \bibinfo{person}{Abhradeep Thakurta}, {and}
  \bibinfo{person}{Oliver Williams}.} \bibinfo{year}{2015}\natexlab{}.
\newblock \showarticletitle{To drop or not to drop: Robustness, consistency and
  differential privacy properties of dropout}.
\newblock \bibinfo{journal}{{\em arXiv:1503.02031\/}} (\bibinfo{year}{2015}).
\newblock


\bibitem[\protect\citeauthoryear{Kasiviswanathan, Lee, Nissim, Raskhodnikova,
  and Smith}{Kasiviswanathan et~al\mbox{.}}{2011}]%
        {kasiviswanathan2011can}
\bibfield{author}{\bibinfo{person}{Shiva~Prasad Kasiviswanathan},
  \bibinfo{person}{Homin~K Lee}, \bibinfo{person}{Kobbi Nissim},
  \bibinfo{person}{Sofya Raskhodnikova}, {and} \bibinfo{person}{Adam Smith}.}
  \bibinfo{year}{2011}\natexlab{}.
\newblock \showarticletitle{What can we learn privately?}
\newblock \bibinfo{journal}{{\it SIAM J. Comput.}} \bibinfo{volume}{40},
  \bibinfo{number}{3} (\bibinfo{year}{2011}), \bibinfo{pages}{793--826}.
\newblock


\bibitem[\protect\citeauthoryear{Kifer and Machanavajjhala}{Kifer and
  Machanavajjhala}{2011}]%
        {kifer2011no}
\bibfield{author}{\bibinfo{person}{Daniel Kifer} {and} \bibinfo{person}{Ashwin
  Machanavajjhala}.} \bibinfo{year}{2011}\natexlab{}.
\newblock \showarticletitle{No free lunch in data privacy}. In
  \bibinfo{booktitle}{{\em Proceedings of the 2011 ACM SIGMOD International
  Conference on Management of data}}. ACM, \bibinfo{pages}{193--204}.
\newblock


\bibitem[\protect\citeauthoryear{Krizhevsky~Alex and Geoffrey}{Krizhevsky~Alex
  and Geoffrey}{[n. d.]}]%
        {cifar10web}
\bibfield{author}{\bibinfo{person}{Nair~Vinod Krizhevsky~Alex} {and}
  \bibinfo{person}{Hinton Geoffrey}.} \bibinfo{year}{[n. d.]}\natexlab{}.
\newblock \bibinfo{title}{CIFAR-10 Dataset}.
\newblock   (\bibinfo{year}{[n. d.]}).
\newblock
\showURL{%
\url{https://www.cs.toronto.edu/~kriz/cifar.html}}


\bibitem[\protect\citeauthoryear{Lai}{Lai}{2015}]%
        {lai2015giraffe}
\bibfield{author}{\bibinfo{person}{Matthew Lai}.}
  \bibinfo{year}{2015}\natexlab{}.
\newblock \showarticletitle{Giraffe: Using deep reinforcement learning to play
  chess}.
\newblock \bibinfo{journal}{{\em arXiv preprint arXiv:1509.01549\/}}
  (\bibinfo{year}{2015}).
\newblock


\bibitem[\protect\citeauthoryear{Lamb, GOYAL, Zhang, Zhang, Courville, and
  Bengio}{Lamb et~al\mbox{.}}{2016}]%
        {Lamb16}
\bibfield{author}{\bibinfo{person}{Alex~M Lamb}, \bibinfo{person}{Anirudh Goyal
  ALIAS~PARTH GOYAL}, \bibinfo{person}{Ying Zhang}, \bibinfo{person}{Saizheng
  Zhang}, \bibinfo{person}{Aaron~C Courville}, {and} \bibinfo{person}{Yoshua
  Bengio}.} \bibinfo{year}{2016}\natexlab{}.
\newblock \showarticletitle{Professor forcing: A new algorithm for training
  recurrent networks}. In \bibinfo{booktitle}{{\em Advances In Neural
  Information Processing Systems}}. \bibinfo{pages}{4601--4609}.
\newblock


\bibitem[\protect\citeauthoryear{Laskov et~al\mbox{.}}{Laskov
  et~al\mbox{.}}{2014}]%
        {laskovPracticalEvasion}
\bibfield{author}{\bibinfo{person}{Pavel Laskov} {et~al\mbox{.}}}
  \bibinfo{year}{2014}\natexlab{}.
\newblock \showarticletitle{Practical evasion of a learning-based classifier: A
  case study}. In \bibinfo{booktitle}{{\em Security and Privacy (SP), 2014 IEEE
  Symposium on}}. IEEE, \bibinfo{pages}{197--211}.
\newblock


\bibitem[\protect\citeauthoryear{LeCun, Bengio, and Hinton}{LeCun
  et~al\mbox{.}}{2015}]%
        {lecun2015deep}
\bibfield{author}{\bibinfo{person}{Yann LeCun}, \bibinfo{person}{Yoshua
  Bengio}, {and} \bibinfo{person}{Geoffrey Hinton}.}
  \bibinfo{year}{2015}\natexlab{}.
\newblock \showarticletitle{Deep learning}.
\newblock \bibinfo{journal}{{\em Nature\/}} \bibinfo{volume}{521},
  \bibinfo{number}{7553} (\bibinfo{year}{2015}), \bibinfo{pages}{436--444}.
\newblock


\bibitem[\protect\citeauthoryear{LeCun, Cortes, and Burges}{LeCun
  et~al\mbox{.}}{1998}]%
        {lecunMnistWeb}
\bibfield{author}{\bibinfo{person}{Yann LeCun}, \bibinfo{person}{Corinna
  Cortes}, {and} \bibinfo{person}{Christopher~J.C. Burges}.}
  \bibinfo{year}{1998}\natexlab{}.
\newblock \bibinfo{title}{The {MNIST} database of handwritten digits}.
\newblock   (\bibinfo{year}{1998}).
\newblock
\showURL{%
\url{http://yann.lecun.com/exdb/mnist/}}


\bibitem[\protect\citeauthoryear{LeCun, Kavukcuoglu, Farabet,
  et~al\mbox{.}}{LeCun et~al\mbox{.}}{2010}]%
        {lecun2010convolutional}
\bibfield{author}{\bibinfo{person}{Yann LeCun}, \bibinfo{person}{Koray
  Kavukcuoglu}, \bibinfo{person}{Cl{\'e}ment Farabet}, {et~al\mbox{.}}}
  \bibinfo{year}{2010}\natexlab{}.
\newblock \showarticletitle{Convolutional networks and applications in
  vision.}. In \bibinfo{booktitle}{{\em ISCAS}}. \bibinfo{pages}{253--256}.
\newblock


\bibitem[\protect\citeauthoryear{Liu, Chakraborty, and Mittal}{Liu
  et~al\mbox{.}}{2016}]%
        {liu2016dependence}
\bibfield{author}{\bibinfo{person}{Changchang Liu}, \bibinfo{person}{Supriyo
  Chakraborty}, {and} \bibinfo{person}{Prateek Mittal}.}
  \bibinfo{year}{2016}\natexlab{}.
\newblock \showarticletitle{Dependence Makes You Vulnerable: Differential
  Privacy Under Dependent Tuples}. In \bibinfo{booktitle}{{\em The Network and
  Distributed System Security Symposium 2016}} {\em (\bibinfo{series}{NDSS
  '16})}. \bibinfo{pages}{1322--1333}.
\newblock
\showURL{%
\url{https://www.internetsociety.org/sites/default/files/blogs-media/dependence-makes-you-vulnerable-differential-privacy-under-dependent-tuples.pdf}}


\bibitem[\protect\citeauthoryear{McCulloch and Pitts}{McCulloch and
  Pitts}{1943}]%
        {McCulloch1943}
\bibfield{author}{\bibinfo{person}{Warren~S. McCulloch} {and}
  \bibinfo{person}{Walter Pitts}.} \bibinfo{year}{1943}\natexlab{}.
\newblock \showarticletitle{A logical calculus of the ideas immanent in nervous
  activity}.
\newblock \bibinfo{journal}{{\em The bulletin of mathematical biophysics\/}}
  \bibinfo{volume}{5}, \bibinfo{number}{4} (\bibinfo{year}{1943}),
  \bibinfo{pages}{115--133}.
\newblock
\showISSN{1522-9602}
\showDOI{%
\url{https://doi.org/10.1007/BF02478259}}


\bibitem[\protect\citeauthoryear{McMahan and Ramage}{McMahan and
  Ramage}{2017}]%
        {brendanRamageFederatedWeb}
\bibfield{author}{\bibinfo{person}{Brendan McMahan} {and}
  \bibinfo{person}{Daniel Ramage}.} \bibinfo{year}{2017}\natexlab{}.
\newblock \bibinfo{title}{Federated Learning: Collaborative Machine Learning
  without Centralized Training Data}.
\newblock   (\bibinfo{year}{2017}).
\newblock
\showURL{%
\url{https://research.googleblog.com/2017/04/federated-learning-collaborative.html}}


\bibitem[\protect\citeauthoryear{McMahan, Moore, Ramage, and Arcas}{McMahan
  et~al\mbox{.}}{2016}]%
        {federatedDeepLearning}
\bibfield{author}{\bibinfo{person}{H.~Brendan McMahan}, \bibinfo{person}{Eider
  Moore}, \bibinfo{person}{Daniel Ramage}, {and} \bibinfo{person}{Blaise
  Ag?era~y Arcas}.} \bibinfo{year}{2016}\natexlab{}.
\newblock \showarticletitle{Federated Learning of Deep Networks using Model
  Averaging}.
\newblock \bibinfo{journal}{{\em arXiv:1502.01710v5\/}} (\bibinfo{year}{2016}).
\newblock


\bibitem[\protect\citeauthoryear{McPherson, Shokri, and Shmatikov}{McPherson
  et~al\mbox{.}}{2016}]%
        {mcpherson2016defeating}
\bibfield{author}{\bibinfo{person}{Richard McPherson}, \bibinfo{person}{Reza
  Shokri}, {and} \bibinfo{person}{Vitaly Shmatikov}.}
  \bibinfo{year}{2016}\natexlab{}.
\newblock \showarticletitle{Defeating Image Obfuscation with Deep Learning}.
\newblock \bibinfo{journal}{{\em arXiv:1609.00408\/}} (\bibinfo{year}{2016}).
\newblock


\bibitem[\protect\citeauthoryear{McSherry}{McSherry}{2016a}]%
        {frankDPcorrelated}
\bibfield{author}{\bibinfo{person}{Frank McSherry}.}
  \bibinfo{year}{2016}\natexlab{a}.
\newblock \bibinfo{title}{Differential Privacy and Correlated Data}.
\newblock   (\bibinfo{year}{2016}).
\newblock
\showURL{%
\url{https://github.com/frankmcsherry/blog/blob/master/posts/2016-08-29.md}}


\bibitem[\protect\citeauthoryear{McSherry}{McSherry}{2016b}]%
        {frankDPLunchtime}
\bibfield{author}{\bibinfo{person}{Frank McSherry}.}
  \bibinfo{year}{2016}\natexlab{b}.
\newblock \bibinfo{title}{Lunchtime for Data Privacy}.
\newblock   (\bibinfo{year}{2016}).
\newblock
\showURL{%
\url{https://github.com/frankmcsherry/blog/blob/master/posts/2016-08-16.md}}


\bibitem[\protect\citeauthoryear{McSherry and Mironov}{McSherry and
  Mironov}{2009}]%
        {mcsherry2009differentially}
\bibfield{author}{\bibinfo{person}{Frank McSherry} {and} \bibinfo{person}{Ilya
  Mironov}.} \bibinfo{year}{2009}\natexlab{}.
\newblock \showarticletitle{Differentially private recommender systems:
  building privacy into the Netflix Prize contenders}. In
  \bibinfo{booktitle}{{\em Proceedings of the 15th ACM SIGKDD international
  conference on Knowledge discovery and data mining}}. ACM,
  \bibinfo{pages}{627--636}.
\newblock


\bibitem[\protect\citeauthoryear{Mescheder, Nowozin, and Geiger}{Mescheder
  et~al\mbox{.}}{2017}]%
        {MeschederNG17}
\bibfield{author}{\bibinfo{person}{Lars Mescheder}, \bibinfo{person}{Sebastian
  Nowozin}, {and} \bibinfo{person}{Andreas Geiger}.}
  \bibinfo{year}{2017}\natexlab{}.
\newblock \showarticletitle{Adversarial Variational Bayes: Unifying Variational
  Autoencoders and Generative Adversarial Networks}.
\newblock \bibinfo{journal}{{\em arXiv preprint arXiv:1701.04722\/}}
  (\bibinfo{year}{2017}).
\newblock


\bibitem[\protect\citeauthoryear{Metz}{Metz}{2016}]%
        {googleGO}
\bibfield{author}{\bibinfo{person}{Cade Metz}.}
  \bibinfo{year}{2016}\natexlab{}.
\newblock \bibinfo{title}{Google's {GO} victory is just a glimpse of how
  powerful ai will be}.
\newblock   (\bibinfo{year}{2016}).
\newblock
\showURL{%
\url{https://tinyurl.com/l6ddhg9}}


\bibitem[\protect\citeauthoryear{Mittal}{Mittal}{2016}]%
        {freedomToTinker}
\bibfield{author}{\bibinfo{person}{Prateek Mittal}.}
  \bibinfo{year}{2016}\natexlab{}.
\newblock \bibinfo{title}{Differential Privacy is Vulnerable to Correlated Data
  {\-} Introducing Dependent Differential Privacy}.
\newblock   (\bibinfo{year}{2016}).
\newblock
\showURL{%
\url{https://tinyurl.com/l3lx7qh}}


\bibitem[\protect\citeauthoryear{Mnih, Badia, Mirza, Graves, Lillicrap, Harley,
  Silver, and Kavukcuoglu}{Mnih et~al\mbox{.}}{2016}]%
        {mnih2016asynchronous}
\bibfield{author}{\bibinfo{person}{Volodymyr Mnih},
  \bibinfo{person}{Adria~Puigdomenech Badia}, \bibinfo{person}{Mehdi Mirza},
  \bibinfo{person}{Alex Graves}, \bibinfo{person}{Timothy~P Lillicrap},
  \bibinfo{person}{Tim Harley}, \bibinfo{person}{David Silver}, {and}
  \bibinfo{person}{Koray Kavukcuoglu}.} \bibinfo{year}{2016}\natexlab{}.
\newblock \showarticletitle{Asynchronous methods for deep reinforcement
  learning}.
\newblock \bibinfo{journal}{{\em arXiv:1602.01783\/}} (\bibinfo{year}{2016}).
\newblock


\bibitem[\protect\citeauthoryear{Mnih, Kavukcuoglu, Silver, Graves, Antonoglou,
  Wierstra, and Riedmiller}{Mnih et~al\mbox{.}}{2013}]%
        {mnih2013playing}
\bibfield{author}{\bibinfo{person}{Volodymyr Mnih}, \bibinfo{person}{Koray
  Kavukcuoglu}, \bibinfo{person}{David Silver}, \bibinfo{person}{Alex Graves},
  \bibinfo{person}{Ioannis Antonoglou}, \bibinfo{person}{Daan Wierstra}, {and}
  \bibinfo{person}{Martin Riedmiller}.} \bibinfo{year}{2013}\natexlab{}.
\newblock \showarticletitle{Playing atari with deep reinforcement learning}.
\newblock \bibinfo{journal}{{\em arXiv:1312.5602\/}} (\bibinfo{year}{2013}).
\newblock


\bibitem[\protect\citeauthoryear{Mohassel and Zhang}{Mohassel and
  Zhang}{2017}]%
        {mohasselsecureml}
\bibfield{author}{\bibinfo{person}{Payman Mohassel} {and}
  \bibinfo{person}{Yupeng Zhang}.} \bibinfo{year}{2017}\natexlab{}.
\newblock \showarticletitle{SecureML: A System for Scalable Privacy-Preserving
  Machine Learning}. In \bibinfo{booktitle}{{\em IEEE Symposium on Security and
  Privacy}}.
\newblock


\bibitem[\protect\citeauthoryear{Narayan, Feldman, Papadimitriou, and
  Haeberlen}{Narayan et~al\mbox{.}}{2015}]%
        {narayan2015verifiable}
\bibfield{author}{\bibinfo{person}{Arjun Narayan}, \bibinfo{person}{Ariel
  Feldman}, \bibinfo{person}{Antonis Papadimitriou}, {and}
  \bibinfo{person}{Andreas Haeberlen}.} \bibinfo{year}{2015}\natexlab{}.
\newblock \showarticletitle{Verifiable differential privacy}. In
  \bibinfo{booktitle}{{\em Proceedings of the Tenth European Conference on
  Computer Systems}}. ACM, \bibinfo{pages}{28}.
\newblock


\bibitem[\protect\citeauthoryear{Ohrimenko, Schuster, Fournet, Mehta, Nowozin,
  Vaswani, and Costa}{Ohrimenko et~al\mbox{.}}{2016}]%
        {ohrimenko2016oblivious}
\bibfield{author}{\bibinfo{person}{Olga Ohrimenko}, \bibinfo{person}{Felix
  Schuster}, \bibinfo{person}{C{\'e}dric Fournet}, \bibinfo{person}{Aastha
  Mehta}, \bibinfo{person}{Sebastian Nowozin}, \bibinfo{person}{Kapil Vaswani},
  {and} \bibinfo{person}{Manuel Costa}.} \bibinfo{year}{2016}\natexlab{}.
\newblock \showarticletitle{Oblivious Multi-Party Machine Learning on Trusted
  Processors}. In \bibinfo{booktitle}{{\em USENIX Security}}.
\newblock


\bibitem[\protect\citeauthoryear{Oord, Dieleman, Zen, Simonyan, Vinyals,
  Graves, Kalchbrenner, Senior, and Kavukcuoglu}{Oord et~al\mbox{.}}{2016}]%
        {oord2016wavenet}
\bibfield{author}{\bibinfo{person}{Aaron van~den Oord}, \bibinfo{person}{Sander
  Dieleman}, \bibinfo{person}{Heiga Zen}, \bibinfo{person}{Karen Simonyan},
  \bibinfo{person}{Oriol Vinyals}, \bibinfo{person}{Alex Graves},
  \bibinfo{person}{Nal Kalchbrenner}, \bibinfo{person}{Andrew Senior}, {and}
  \bibinfo{person}{Koray Kavukcuoglu}.} \bibinfo{year}{2016}\natexlab{}.
\newblock \showarticletitle{WaveNet: A generative model for raw audio}.
\newblock \bibinfo{journal}{{\em arXiv:1609.03499\/}} (\bibinfo{year}{2016}).
\newblock


\bibitem[\protect\citeauthoryear{Papernot, McDaniel, and Goodfellow}{Papernot
  et~al\mbox{.}}{2016}]%
        {papernot2016transferability}
\bibfield{author}{\bibinfo{person}{Nicolas Papernot}, \bibinfo{person}{Patrick
  McDaniel}, {and} \bibinfo{person}{Ian Goodfellow}.}
  \bibinfo{year}{2016}\natexlab{}.
\newblock \showarticletitle{Transferability in Machine Learning: from Phenomena
  to Black-Box Attacks using Adversarial Samples}.
\newblock \bibinfo{journal}{{\em arXiv preprint arXiv:1605.07277\/}}
  (\bibinfo{year}{2016}).
\newblock


\bibitem[\protect\citeauthoryear{Papernot, McDaniel, Jha, Fredrikson, Celik,
  and Swami}{Papernot et~al\mbox{.}}{2015}]%
        {papernot2015limitations}
\bibfield{author}{\bibinfo{person}{Nicolas Papernot}, \bibinfo{person}{Patrick
  McDaniel}, \bibinfo{person}{Somesh Jha}, \bibinfo{person}{Matt Fredrikson},
  \bibinfo{person}{Z~Berkay Celik}, {and} \bibinfo{person}{Ananthram Swami}.}
  \bibinfo{year}{2015}\natexlab{}.
\newblock \showarticletitle{The Limitations of Deep Learning in Adversarial
  Settings}.
\newblock \bibinfo{journal}{{\em Proceedings of the 1st IEEE European Symposium
  on Security and Privacy\/}} (\bibinfo{year}{2015}).
\newblock


\bibitem[\protect\citeauthoryear{Pathak, Rane, and Raj}{Pathak
  et~al\mbox{.}}{2010}]%
        {pathak2010multiparty}
\bibfield{author}{\bibinfo{person}{Manas Pathak}, \bibinfo{person}{Shantanu
  Rane}, {and} \bibinfo{person}{Bhiksha Raj}.} \bibinfo{year}{2010}\natexlab{}.
\newblock \showarticletitle{Multiparty differential privacy via aggregation of
  locally trained classifiers}. In \bibinfo{booktitle}{{\em Advances in Neural
  Information Processing Systems}}. \bibinfo{pages}{1876--1884}.
\newblock


\bibitem[\protect\citeauthoryear{Perarnau, van~de Weijer, Raducanu, and
  {\'A}lvarez}{Perarnau et~al\mbox{.}}{2016}]%
        {perarnau2016invertible}
\bibfield{author}{\bibinfo{person}{Guim Perarnau}, \bibinfo{person}{Joost
  van~de Weijer}, \bibinfo{person}{Bogdan Raducanu}, {and}
  \bibinfo{person}{Jose~M {\'A}lvarez}.} \bibinfo{year}{2016}\natexlab{}.
\newblock \showarticletitle{Invertible Conditional GANs for image editing}.
\newblock \bibinfo{journal}{{\em arXiv preprint arXiv:1611.06355\/}}
  (\bibinfo{year}{2016}).
\newblock


\bibitem[\protect\citeauthoryear{Phan, Wang, Wu, and Dou}{Phan
  et~al\mbox{.}}{2016}]%
        {phan2016differential}
\bibfield{author}{\bibinfo{person}{NhatHai Phan}, \bibinfo{person}{Yue Wang},
  \bibinfo{person}{Xintao Wu}, {and} \bibinfo{person}{Dejing Dou}.}
  \bibinfo{year}{2016}\natexlab{}.
\newblock \showarticletitle{Differential Privacy Preservation for Deep
  Auto-Encoders: an Application of Human Behavior Prediction}. In
  \bibinfo{booktitle}{{\em Proceedings of the 30th AAAI Conference on
  Artificial Intelligence, AAAI}}. \bibinfo{pages}{12--17}.
\newblock


\bibitem[\protect\citeauthoryear{Radford, Metz, and Chintala}{Radford
  et~al\mbox{.}}{2016}]%
        {Radford16}
\bibfield{author}{\bibinfo{person}{Alec Radford}, \bibinfo{person}{Luke Metz},
  {and} \bibinfo{person}{Soumith Chintala}.} \bibinfo{year}{2016}\natexlab{}.
\newblock \showarticletitle{Unsupervised Representation Learning with Deep
  Convolutional Generative Adversarial Networks}. In \bibinfo{booktitle}{{\em
  4th International Conference on Learning Representations}}.
\newblock


\bibitem[\protect\citeauthoryear{Rasmussen and Williams}{Rasmussen and
  Williams}{2006}]%
        {Rasmussen06}
\bibfield{author}{\bibinfo{person}{C.~E. Rasmussen} {and}
  \bibinfo{person}{C.~K.~I. Williams}.} \bibinfo{year}{2006}\natexlab{}.
\newblock \bibinfo{booktitle}{{\em Gaussian Processes for Machine Learning}}.
\newblock \bibinfo{publisher}{MIT Press}, \bibinfo{address}{Cambridge, MA}.
\newblock


\bibitem[\protect\citeauthoryear{Salimans, Goodfellow, Zaremba, Cheung,
  Radford, and Chen}{Salimans et~al\mbox{.}}{2016}]%
        {Salimans16}
\bibfield{author}{\bibinfo{person}{Tim Salimans}, \bibinfo{person}{Ian
  Goodfellow}, \bibinfo{person}{Wojciech Zaremba}, \bibinfo{person}{Vicki
  Cheung}, \bibinfo{person}{Alec Radford}, {and} \bibinfo{person}{Xi Chen}.}
  \bibinfo{year}{2016}\natexlab{}.
\newblock \showarticletitle{Improved techniques for training gans}. In
  \bibinfo{booktitle}{{\em Advances in Neural Information Processing Systems}}.
  \bibinfo{pages}{2226--2234}.
\newblock


\bibitem[\protect\citeauthoryear{Samaria and Harter}{Samaria and
  Harter}{1994}]%
        {samaria1994parameterisation}
\bibfield{author}{\bibinfo{person}{Ferdinando~S Samaria} {and}
  \bibinfo{person}{Andy~C Harter}.} \bibinfo{year}{1994}\natexlab{}.
\newblock \showarticletitle{Parameterisation of a stochastic model for human
  face identification}. In \bibinfo{booktitle}{{\em Applications of Computer
  Vision, 1994., Proceedings of the Second IEEE Workshop on}}. IEEE,
  \bibinfo{pages}{138--142}.
\newblock


\bibitem[\protect\citeauthoryear{Sarwate and Chaudhuri}{Sarwate and
  Chaudhuri}{2013}]%
        {sarwate2013signal}
\bibfield{author}{\bibinfo{person}{Anand~D Sarwate} {and}
  \bibinfo{person}{Kamalika Chaudhuri}.} \bibinfo{year}{2013}\natexlab{}.
\newblock \showarticletitle{Signal processing and machine learning with
  differential privacy: Algorithms and challenges for continuous data}.
\newblock \bibinfo{journal}{{\em IEEE signal processing magazine\/}}
  \bibinfo{volume}{30}, \bibinfo{number}{5} (\bibinfo{year}{2013}),
  \bibinfo{pages}{86--94}.
\newblock


\bibitem[\protect\citeauthoryear{Schmidhuber}{Schmidhuber}{2015}]%
        {Schmidhuber201585}
\bibfield{author}{\bibinfo{person}{J{\"u}rgen Schmidhuber}.}
  \bibinfo{year}{2015}\natexlab{}.
\newblock \showarticletitle{Deep learning in neural networks: An overview}.
\newblock \bibinfo{journal}{{\em Neural networks\/}}  \bibinfo{volume}{61}
  (\bibinfo{year}{2015}), \bibinfo{pages}{85--117}.
\newblock


\bibitem[\protect\citeauthoryear{Scholkopf and Smola}{Scholkopf and
  Smola}{2001}]%
        {Schoelkopf02}
\bibfield{author}{\bibinfo{person}{Bernhard Scholkopf} {and}
  \bibinfo{person}{Alexander~J Smola}.} \bibinfo{year}{2001}\natexlab{}.
\newblock \bibinfo{booktitle}{{\em Learning with kernels: support vector
  machines, regularization, optimization, and beyond}}.
\newblock \bibinfo{publisher}{MIT press}.
\newblock


\bibitem[\protect\citeauthoryear{Shokri and Shmatikov}{Shokri and
  Shmatikov}{2015}]%
        {shokriPPDL}
\bibfield{author}{\bibinfo{person}{Reza Shokri} {and} \bibinfo{person}{Vitaly
  Shmatikov}.} \bibinfo{year}{2015}\natexlab{}.
\newblock \showarticletitle{Privacy-Preserving Deep Learning}. In
  \bibinfo{booktitle}{{\em Proceedings of the 22Nd ACM SIGSAC Conference on
  Computer and Communications Security}} {\em (\bibinfo{series}{CCS '15})}.
  \bibinfo{publisher}{ACM}, \bibinfo{pages}{1310--1321}.
\newblock
\showISBNx{978-1-4503-3832-5}
\showDOI{%
\url{https://doi.org/10.1145/2810103.2813687}}


\bibitem[\protect\citeauthoryear{Shokri, Stronati, Song, and Shmatikov}{Shokri
  et~al\mbox{.}}{2017}]%
        {shokri2017membership}
\bibfield{author}{\bibinfo{person}{Reza Shokri}, \bibinfo{person}{Marco
  Stronati}, \bibinfo{person}{Congzheng Song}, {and} \bibinfo{person}{Vitaly
  Shmatikov}.} \bibinfo{year}{2017}\natexlab{}.
\newblock \showarticletitle{Membership Inference Attacks against Machine
  Learning Models}. In \bibinfo{booktitle}{{\em IEEE Symposium on Security and
  Privacy (S\&P), Oakland}}.
\newblock


\bibitem[\protect\citeauthoryear{Song, Chaudhuri, and Sarwate}{Song
  et~al\mbox{.}}{2013}]%
        {song2013stochastic}
\bibfield{author}{\bibinfo{person}{Shuang Song}, \bibinfo{person}{Kamalika
  Chaudhuri}, {and} \bibinfo{person}{Anand~D Sarwate}.}
  \bibinfo{year}{2013}\natexlab{}.
\newblock \showarticletitle{Stochastic gradient descent with differentially
  private updates}. In \bibinfo{booktitle}{{\em Global Conference on Signal and
  Information Processing (GlobalSIP), 2013 IEEE}}. IEEE,
  \bibinfo{pages}{245--248}.
\newblock


\bibitem[\protect\citeauthoryear{Srinivasan, Fearon, Alcicek, Nair, Blackwell,
  Panneershelvam, De~Maria, Mnih, Kavukcuoglu, Silver,
  et~al\mbox{.}}{Srinivasan et~al\mbox{.}}{2016}]%
        {srinivasan2016distributed}
\bibfield{author}{\bibinfo{person}{Praveen~Deepak Srinivasan},
  \bibinfo{person}{Rory Fearon}, \bibinfo{person}{Cagdas Alcicek},
  \bibinfo{person}{Arun~Sarath Nair}, \bibinfo{person}{Samuel Blackwell},
  \bibinfo{person}{Vedavyas Panneershelvam}, \bibinfo{person}{Alessandro
  De~Maria}, \bibinfo{person}{Volodymyr Mnih}, \bibinfo{person}{Koray
  Kavukcuoglu}, \bibinfo{person}{David Silver}, {et~al\mbox{.}}}
  \bibinfo{year}{2016}\natexlab{}.
\newblock \bibinfo{title}{Distributed training of reinforcement learning
  systems}.
\newblock   (\bibinfo{date}{Feb.~4} \bibinfo{year}{2016}).
\newblock
\newblock
\shownote{US Patent App. 15/016,173.}


\bibitem[\protect\citeauthoryear{Szegedy, Zaremba, Sutskever, Bruna, Erhan,
  Goodfellow, and Fergus}{Szegedy et~al\mbox{.}}{2014}]%
        {42503}
\bibfield{author}{\bibinfo{person}{Christian Szegedy},
  \bibinfo{person}{Wojciech Zaremba}, \bibinfo{person}{Ilya Sutskever},
  \bibinfo{person}{Joan Bruna}, \bibinfo{person}{Dumitru Erhan},
  \bibinfo{person}{Ian Goodfellow}, {and} \bibinfo{person}{Rob Fergus}.}
  \bibinfo{year}{2014}\natexlab{}.
\newblock \showarticletitle{Intriguing properties of neural networks}. In
  \bibinfo{booktitle}{{\em International Conference on Learning
  Representations}}.
\newblock
\showURL{%
\url{http://arxiv.org/abs/1312.6199}}


\bibitem[\protect\citeauthoryear{Taigman, Yang, Ranzato, and Wolf}{Taigman
  et~al\mbox{.}}{2014}]%
        {Taigman:2014:DCG:2679600.2680208}
\bibfield{author}{\bibinfo{person}{Yaniv Taigman}, \bibinfo{person}{Ming Yang},
  \bibinfo{person}{Marc'Aurelio Ranzato}, {and} \bibinfo{person}{Lior Wolf}.}
  \bibinfo{year}{2014}\natexlab{}.
\newblock \showarticletitle{DeepFace: Closing the Gap to Human-Level
  Performance in Face Verification}. In \bibinfo{booktitle}{{\em Proceedings of
  the 2014 IEEE Conference on Computer Vision and Pattern Recognition}} {\em
  (\bibinfo{series}{CVPR '14})}. \bibinfo{publisher}{IEEE Computer Society},
  \bibinfo{address}{Washington, DC, USA}, \bibinfo{pages}{1701--1708}.
\newblock
\showISBNx{978-1-4799-5118-5}
\showDOI{%
\url{https://doi.org/10.1109/CVPR.2014.220}}


\bibitem[\protect\citeauthoryear{Tram{\`e}r, Zhang, Juels, Reiter, and
  Ristenpart}{Tram{\`e}r et~al\mbox{.}}{2016}]%
        {tramer2016stealing}
\bibfield{author}{\bibinfo{person}{Florian Tram{\`e}r}, \bibinfo{person}{Fan
  Zhang}, \bibinfo{person}{Ari Juels}, \bibinfo{person}{Michael~K Reiter},
  {and} \bibinfo{person}{Thomas Ristenpart}.} \bibinfo{year}{2016}\natexlab{}.
\newblock \showarticletitle{Stealing Machine Learning Models via Prediction
  APIs}. In \bibinfo{booktitle}{{\em USENIX Security}}.
\newblock


\bibitem[\protect\citeauthoryear{Vapnik and Vapnik}{Vapnik and Vapnik}{1998}]%
        {Vapnik98}
\bibfield{author}{\bibinfo{person}{Vladimir~Naumovich Vapnik} {and}
  \bibinfo{person}{Vlamimir Vapnik}.} \bibinfo{year}{1998}\natexlab{}.
\newblock \bibinfo{booktitle}{{\em Statistical learning theory}}.
  Vol.~\bibinfo{volume}{1}.
\newblock \bibinfo{publisher}{Wiley New York}.
\newblock


\bibitem[\protect\citeauthoryear{Wainwright, Jordan, and Duchi}{Wainwright
  et~al\mbox{.}}{2012}]%
        {wainwright2012privacy}
\bibfield{author}{\bibinfo{person}{Martin~J Wainwright},
  \bibinfo{person}{Michael~I Jordan}, {and} \bibinfo{person}{John~C Duchi}.}
  \bibinfo{year}{2012}\natexlab{}.
\newblock \showarticletitle{Privacy aware learning}. In
  \bibinfo{booktitle}{{\em Advances in Neural Information Processing Systems}}.
  \bibinfo{pages}{1430--1438}.
\newblock


\bibitem[\protect\citeauthoryear{Xie, Bilenko, Finley, Gilad-Bachrach, Lauter,
  and Naehrig}{Xie et~al\mbox{.}}{2014}]%
        {xie2014crypto}
\bibfield{author}{\bibinfo{person}{Pengtao Xie}, \bibinfo{person}{Misha
  Bilenko}, \bibinfo{person}{Tom Finley}, \bibinfo{person}{Ran Gilad-Bachrach},
  \bibinfo{person}{Kristin Lauter}, {and} \bibinfo{person}{Michael Naehrig}.}
  \bibinfo{year}{2014}\natexlab{}.
\newblock \showarticletitle{Crypto-nets: Neural networks over encrypted data}.
\newblock \bibinfo{journal}{{\em arXiv preprint arXiv:1412.6181\/}}
  (\bibinfo{year}{2014}).
\newblock


\bibitem[\protect\citeauthoryear{Xu, Qi, and Evans}{Xu et~al\mbox{.}}{2016}]%
        {xu2016automatically}
\bibfield{author}{\bibinfo{person}{Weilin Xu}, \bibinfo{person}{Yanjun Qi},
  {and} \bibinfo{person}{David Evans}.} \bibinfo{year}{2016}\natexlab{}.
\newblock \showarticletitle{Automatically evading classifiers}. In
  \bibinfo{booktitle}{{\em NDSS'16}}.
\newblock


\bibitem[\protect\citeauthoryear{Zhang, Zhang, Xiao, Yang, and Winslett}{Zhang
  et~al\mbox{.}}{2012}]%
        {zhang2012functional}
\bibfield{author}{\bibinfo{person}{Jun Zhang}, \bibinfo{person}{Zhenjie Zhang},
  \bibinfo{person}{Xiaokui Xiao}, \bibinfo{person}{Yin Yang}, {and}
  \bibinfo{person}{Marianne Winslett}.} \bibinfo{year}{2012}\natexlab{}.
\newblock \showarticletitle{Functional mechanism: regression analysis under
  differential privacy}.
\newblock \bibinfo{journal}{{\em Proceedings of the VLDB Endowment\/}}
  \bibinfo{volume}{5}, \bibinfo{number}{11} (\bibinfo{year}{2012}),
  \bibinfo{pages}{1364--1375}.
\newblock


\bibitem[\protect\citeauthoryear{Zhang}{Zhang}{2004}]%
        {zhang2004solving}
\bibfield{author}{\bibinfo{person}{Tong Zhang}.}
  \bibinfo{year}{2004}\natexlab{}.
\newblock \showarticletitle{Solving large scale linear prediction problems
  using stochastic gradient descent algorithms}. In \bibinfo{booktitle}{{\em
  Proceedings of the twenty-first international conference on Machine
  learning}}. ACM, \bibinfo{pages}{116}.
\newblock


\bibitem[\protect\citeauthoryear{Zhang and LeCun}{Zhang and LeCun}{2016}]%
        {leCunnNLP}
\bibfield{author}{\bibinfo{person}{Xiang Zhang} {and}
  \bibinfo{person}{Yann~Andr{\'e} LeCun}.} \bibinfo{year}{2016}\natexlab{}.
\newblock \showarticletitle{Text Understanding from Scratch}.
\newblock \bibinfo{journal}{{\em arXiv preprint arXiv:1502.01710v5\/}}
  (\bibinfo{year}{2016}).
\newblock


\bibitem[\protect\citeauthoryear{Zinkevich, Weimer, Li, and Smola}{Zinkevich
  et~al\mbox{.}}{2010}]%
        {zinkevich2010parallelized}
\bibfield{author}{\bibinfo{person}{Martin Zinkevich}, \bibinfo{person}{Markus
  Weimer}, \bibinfo{person}{Lihong Li}, {and} \bibinfo{person}{Alex~J Smola}.}
  \bibinfo{year}{2010}\natexlab{}.
\newblock \showarticletitle{Parallelized stochastic gradient descent}. In
  \bibinfo{booktitle}{{\em Advances in neural information processing systems}}.
  \bibinfo{pages}{2595--2603}.
\newblock


\end{thebibliography}

\vfill\eject
\onecolumn
\appendix
\section{System Architecture}
\label{sysarch}

\begin{figure}[h]
\centering
\includegraphics[width=4.7in]{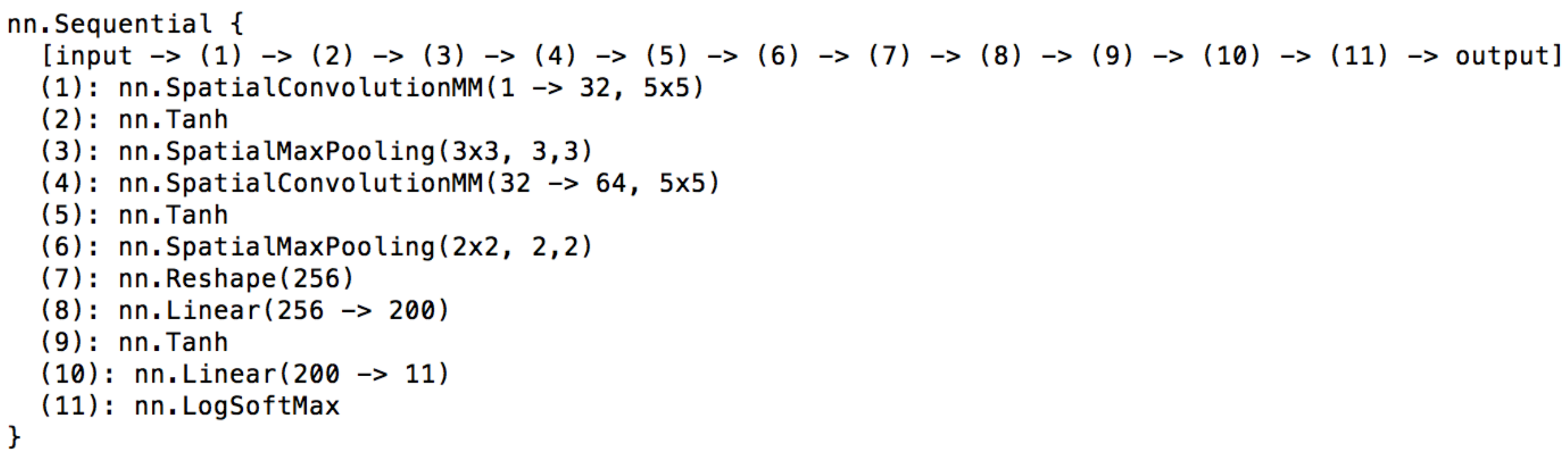}
\caption{Convolutional Neural Network Architecture used for MNIST related experiments, as printed by Torch. Note that the same architecture is used for both the collaboratively trained model and the local discriminator (D) model used by the Adversary}
\label{cnn_architecture_mnist}
\end{figure}

\begin{figure}[h]
\centering
\includegraphics[width=4.7in]{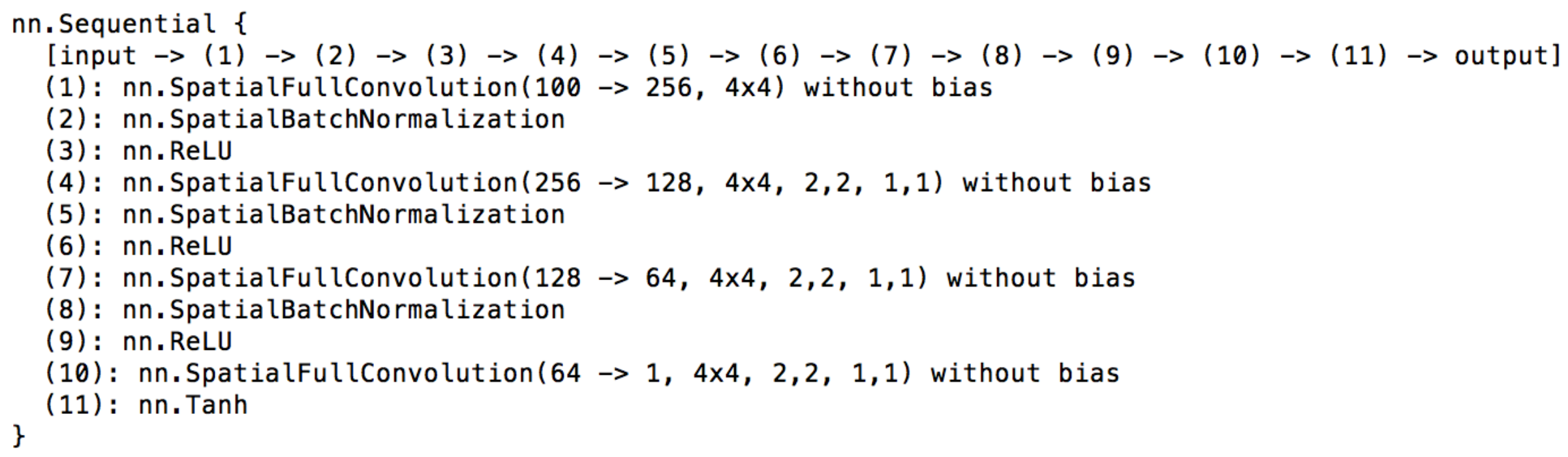}
\caption{Generator Model Architecture used in MNIST experiments}
\label{g_architecture}
\end{figure}

\begin{figure}[h]
  \includegraphics[width=5in]{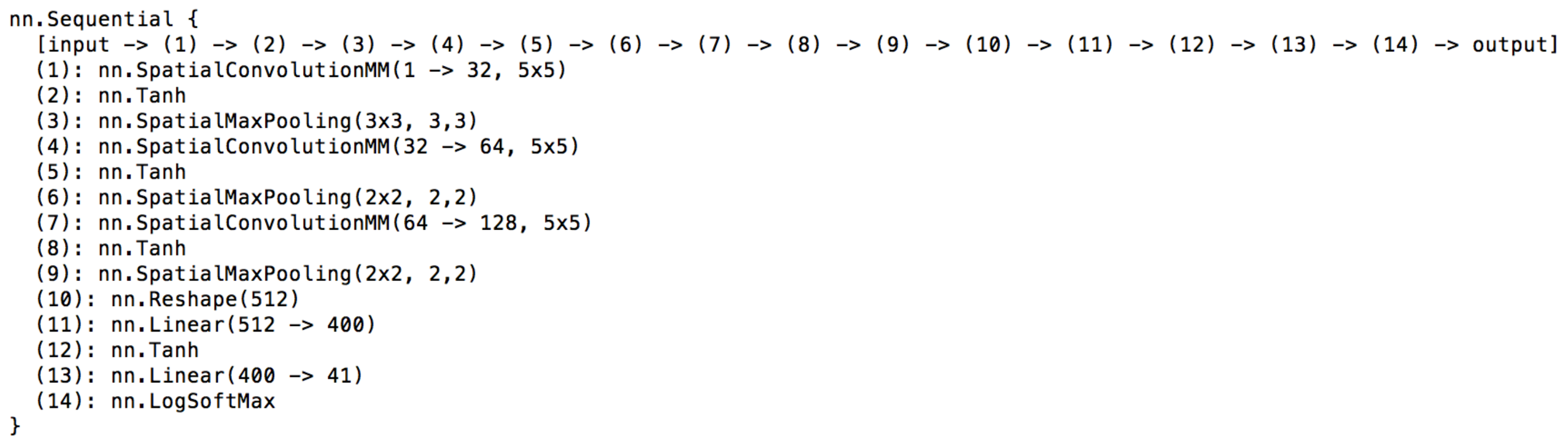}
  \caption{Architecture of the Collaborative Model and the Discriminator (D) utilized in AT\&T Dataset related experiments}
  \label{cnn_architecture_att}
\end{figure}

\begin{figure}[h]
  \includegraphics[width=5in]{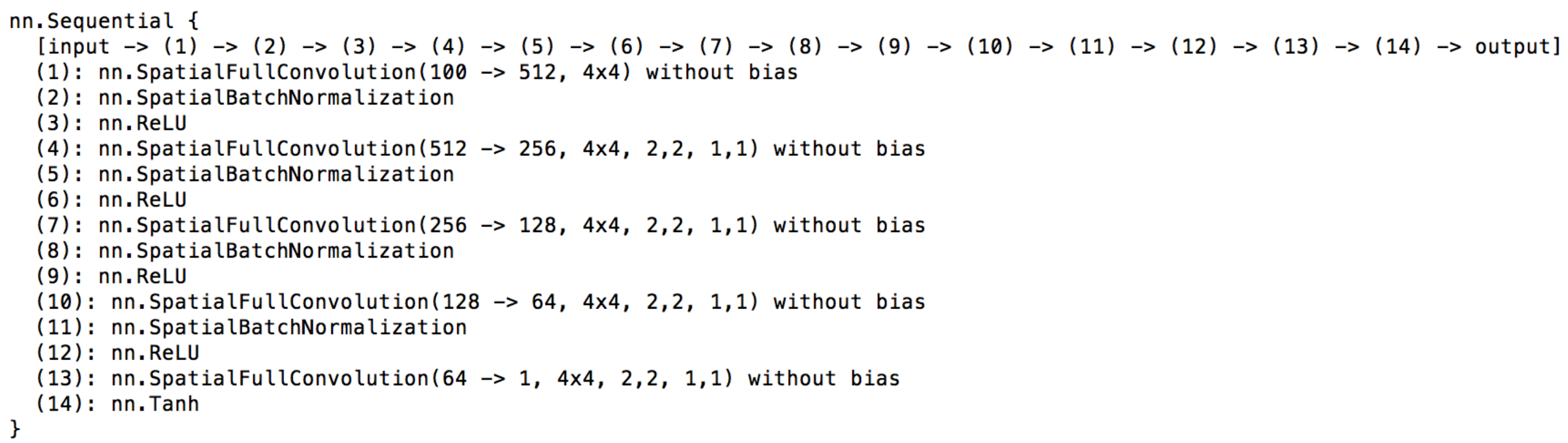}
  \caption{Generator (G) Architecture used in AT\&T Dataset related experiments, as printed by Torch7}
  \label{fig:dcgan_faces_generator}
\end{figure}

\end{document}